DEEP LEARNING AIDED PARAMETRIC CHANNEL COVARIANCE
MATRIX ESTIMATION FOR MILLIMETER WAVE HYBRID MASSIVE
MIMO

A THESIS SUBMITTED TO
THE GRADUATE SCHOOL OF NATURAL AND APPLIED SCIENCES
OF
MIDDLE EAST TECHNICAL UNIVERSITY

BY

ESEN ÖZBAY

IN PARTIAL FULFILLMENT OF THE REQUIREMENTS
FOR
THE DEGREE OF MASTER OF SCIENCE
IN
ELECTRICAL AND ELECTRONIC ENGINEERING

SEPTEMBER 2021

Approval of the thesis:

**DEEP LEARNING AIDED PARAMETRIC CHANNEL COVARIANCE MATRIX ESTIMATION FOR MILLIMETER WAVE HYBRID MASSIVE MIMO**

submitted by **ESEN ÖZBAY** in partial fulfillment of the requirements for the degree of **Master of Science** i**n Electrical and Electronic Engineering, Middle East Technical University** by,

Prof. Dr. Halil Kalıpçılar
Dean, Graduate School of **Natural and Applied Sciences** \_\_\_\_\_\_\_\_\_\_\_\_\_\_\_\_

Prof. Dr. İlkay Ulusoy
Head of the Department, **Electrical and Electronics Engineeering** \_\_\_\_\_\_\_\_\_\_\_\_\_\_\_\_

Assist. Prof. Dr. Gökhan Muzaffer Güvensen
Supervisor, **Electrical and Electronics Engineering, METU** \_\_\_\_\_\_\_\_\_\_\_\_\_\_\_\_

**Examining Committee Members:**

Prof. Dr. Ali Özgür Yılmaz
Electrical and Electronics Eng., METU \_\_\_\_\_\_\_\_\_\_\_\_\_\_\_\_

Assist. Prof. Dr. Gökhan Muzaffer Güvensen
Electrical and Electronics Eng., METU \_\_\_\_\_\_\_\_\_\_\_\_\_\_\_\_

Assoc. Prof. Dr. Tolga Girici
Electrical and Electronics Eng., TOBB University of Economics
and Technology \_\_\_\_\_\_\_\_\_\_\_\_\_\_\_\_

Assist. Prof. Dr. Elif Tuğçe Ceran Arslan
Electrical and Electronics Eng., METU \_\_\_\_\_\_\_\_\_\_\_\_\_\_\_\_

Assist. Prof. Dr. Barış Yüksekkaya
Electrical and Electronics Eng., Hacettepe University \_\_\_\_\_\_\_\_\_\_\_\_\_\_\_\_

Date: 09.09.2021

**I hereby declare that all information in this document has been obtained and presented in accordance with academic rules and ethical conduct. I also declare that, as required by these rules and conduct, I have fully cited and referenced all material and results that are not original to this work.**

Name Last name :

Signature :

# ABSTRACT

## DEEP LEARNING AIDED PARAMETRIC CHANNEL COVARIANCE MATRIX ESTIMATION FOR MILLIMETER WAVE HYBRID MASSIVE MIMO

Özbay, Esen
Master of Science, Electrical and Electronic Engineering
Supervisor : Assist. Prof. Dr. Gökhan Muzaffer Güvensen

Millimeter-wave (mmWave) channels, which occupy frequency ranges much higher than those being used in previous wireless communications systems, are utilized to meet the increased throughput requirements that come with 5G communications. The high levels of attenuation experienced by electromagnetic waves in these frequencies causes MIMO channels to have high spatial correlation. To attain desirable error performances, systems require knowledge about the channel correlations. In this thesis, a deep neural network aided method is proposed for the parametric estimation of the channel covariance matrix (CCM), which contains information regarding the channel correlations. When compared to some methods found in the literature, the proposed method yields satisfactory peformance in terms of both computational complexity and channel estimation errors.

# ÖZ

## MİLİMETRE DALGA HİBRİT KİTLESEL MIMO İÇİN DERİN ÖĞRENME DESTEKLİ PARAMETRİK KANAL KOVARYANS MATRİSİ KESTİRİMİ

Özbay, Esen
Yüksek Lisans, Elektrik ve Elektronik Mühendisliği
Tez Yöneticisi: Dr. Öğr. Üy. Gökhan Muzaffer Güvensen

5G sistemlerde artan iletim hızı ihtiyaçlarını karşılamak için, önceden kullanılanlara kıyasla çok yüksek frekanslarda bulunan milimetre-dalga (mmWave) bantları kullanılmaktadır. Bu frekans bantlarında dalga zayıflaması yüksek olduğundan, çok-girdili-çok-çıktılı (MIMO) sistemlerde uzaysal korelasyonu yüksek kanallar meydana gelmektedir. İstenen başarımların elde edilebilmesi için ise kanal korrelasyonlarının sistem tarafından bilinmesi gereklidir. Bu tezde seyrek kanallar için kanal korelasyon bilgisini içeren kanal kovaryans matrisinin (CCM) parametrik olarak kestirimi için yapay sinir ağları destekli bir yöntem önerilmiştir. Önerilen yapay sinir ağları destekli yöntem, hem hesaplama karmaşıklığı açısından hem de kestirim hatası açısından referans yöntemlere kıyasla tatmin edici başarım sergilemiştir.

Anahtar Kelimeler: MIMO, hibrit hüzmeleme, CCM, 5G, milimetre-dalga

To every graduate student who dreads writing their thesis.

# ACKNOWLEDGMENTS

I would like to thank my supervisor Asst. Prof. Dr. Gökhan Muzaffer Güvensen for his guidance throughout the research process.

I would also like to thank Prof. Dr. Ali Özgür Yılmaz for his guidance during the early stages of my graduate studies.

I would like to thank my father, Prof. Dr. Ekmel Özbay, and my mother, Prof. Dr. Deniz Üner for the academic vision and advice that has enabled me to pursue and complete this degree. I would also like to thank my siblings, İmre Özbay, Fırat Özbay, Sedef Özbay, Umut Yaşar Özbay and Timur Özbay.

I would like to express my gratitude to my friends Dr. Özlem Tuğfe Demir and Utku Çelebi. The emotional support and the scientific advice that they have provided have greatly helped me in writing this thesis. I also want to sincerely thank my dear friends Hanife Akdoğanbulut, Gökcem Tonyalı Karslı and Hakan Karslı for always being there to support me without hesitation. I would also like to thank my friends Esma Ceren Kalfa, Fatma Atıntaş, Alper Şahin and Aslı Soydemir for their companionship as I finished writing this thesis.

This work was financially supported by Turkcell Teknoloji A.Ş., under the '5G and Beyond Graduate Support Program', organized by the Information and Communication Technologies Authority of the Republic of Turkey.

**TABLE OF CONTENTS**

# LIST OF TABLES

TABLES

# LIST OF FIGURES

# LIST OF ABBREVIATIONS

ABBREVIATIONS

| | |
|---|---|
| ANN | Artificial neural network |
| AoA | Angle of arrival |
| AS | Angular spread |
| AWGN | Additive white gaussian noise |
| BF | Beamforming |
| BPSK | Binary Phase Shift Keying |
| BS | Base station |
| CCM | Channel covariance catrix |
| CDF | Cumulative distribution function |
| DBF | Digital beamforming |
| DFT | Disctrete Fourier Transform |
| DNN | Deep neural network |
| EE | Energy efficiency |
| eMBB | Enhanced Mobile Broadband |
| FDD | Frequency division duplex |
| FFT | Fast Fourier Transform |
| HBF | Hybrid beamforming |
| IoT | Internet of Things |
| LTE | Long term evolution |
| MIMO | Multi-input multi-output |
| mMIMO | Massive MIMO |
| mMTC | Massive Machine-Type Communications |
| mmWave | Millimeter-wave |
| MSE | Mean squared error |
| MU-MIMO | Multi-user MIMO |
| MUSIC | Multiple Signal Classification Algorithm |
| OFDM | Orthogonal frequency-division multiplexing |
| OFDMA | Orthogonal frequency-division multiple access |
| PAS | Power angular spectrum. |
| PDF | Probability distribution function |
| PMF | Probability mass function |
| QAM | Quadrature Amplitude Modulation |
| QPSK | Quadrature Phase Shift Keying |
| RF | Radio frequency |
| SE | Spectral efficiency |
| SIMO | Single-input multi-output |
| SISO | Single-input single-output |
| TDD | Time division duplex |
| ULA | Uniform linear array |
| URLLC | Ultra Reliable Low Latency Communication |

# LIST OF SYMBOLS

SYMBOLS

| | |
|---|---|
| $\theta_k^l$ | The AoA of the $l$th active tap of the $k$th user |
| $\sigma_k^l$ | The beam spread of the $l$th active tap of the $k$th user |
| $\rho_k^l$ | The received SNR of the $l$th active tap of the $k$th user |
| $\tau_k^l$ | The delay of the $l$th active tap of the $k$th user |
| $\Phi$ | $N \times 8$ matrix containing DFT beams correponding to angles of interest |
| $\lambda$ | Carrier wavelength |
| $\boldsymbol{b}_k$ | Dimension-reduced received signal for the $k-$th user |
| $c$ | The speed of light, $3 \times 10^8 \, m/s$ |
| $f_c$ | Center frequency |
| $\boldsymbol{h}_k^l$ | $N \times 1$ SIMO channel experienced by the $l$th active tap of the $k$th user |
| $k$ | User index |
| $K$ | Number of single-antenna users |
| $l$ | Active channel tap index for a given user |
| $L$ | Channel delay spread in terms of channel taps |
| $L_{total}$ | The total number of active user taps, $L_{total} = \sum_{k=1}^{K} L_k$. |
| $L_k$ | The number of active taps for each user |
| $N$ | Number of base station antennae |
| $N_{sec}$ | Number of DFT beams per sector for AoA estimation |
| $N_{sec,2}$ | Number of DFT beams per sector for AS estimation |
| $\boldsymbol{R}_k^l$ | $N \times N$ channel covariance of the channel experienced by the $l$th active tap of the $k$th user |
| $T$ | Number of training symbols used for CCM estimation |
| $T_r$ | Number of AoA estimates performed |
| $\boldsymbol{x}_k$ | $1 \times T$ pilot sequence for the $k$-th user |
| $Y$ | $N \times T$ received signal matrix |

# CHAPTER 1

# INTRODUCTION

5G technologies have been commercially available since 2019. This new generation of mobile communications aims to fulfill modern demands for speed and versatility by introducing some major paradigm shifts in terms of network operation, while also preserving many existing architectures.

Two major differences between 5G and the previous generations is the utilization of the millimiter wave (mmWave) frequency band, which has significantly different characteristics than the previously used lower frequency bands, and the use of large scale antenna arrays, which basically adds a new dimension to the system that needs to be processed.

Because the attenuation of electromagnetic waves happens much more rapidly at higher frequencies, mmWave bands are characterized by a lower number of scatterers, resulting in more correlated channels for a given antenna geometry and transmission medium [1]. In Sub-6GHz bands, where there are typically many scatterers in the transmission medium, it is generally assumed that MIMO channels are spatially white, i.e., that the channels of different antennae are independently distributed[1] [2]. This assumption is not valid for mmWave channels, because even if the medium is physically the same, waves at mmWave frequencies usually cannot reach the receivers after being scattered by more than one or two scatterers.

Since the channel distributions of the antennae are not spatially white, a new channel characteritic, namely channel covariance matrices, come into play. Channel covariance matrices (CCMs), when paired with the individual distribution of each channel, completely define the spatial statistics of a mmWave MIMO channel. The distributions of individual channels are generally the same as in Sub-6GHz bands, such as Rayleigh or Rician distributions.

---

[1] The correctness of this assumption depends on the antenna geometry. The antennae must have a separation of at least half the carrier wavelength in order for this assumption to hold. Since antenna arrays are generally designed according to these limitations, this is a safe assumption to make.

CCMs find many uses in mmWave communications. One such use is dimension reduction, which is practically necessary for 5G applications where massive MIMO is used. The large number of transmitting and receiving antennas in massive MIMO systems makes the computational complexity of many elementary operations infeasible without dimension reduction [3]. CCMs can also be used to overcome other challenges of massive MIMO, such as prohibitive computational complexity, uplink pilot contamination, and training overhead [4], [5].

Other than dimension reduction, the statistical channel information contained in CCMs can be used to improve system performance in a variety of ways. One such application is beamforming [6]. Since the basic purpose of beamforming is to create isolated virtual spatial channels, the statistical information contained in CCMs is essential to the beamforming operation.

Channel covariances depend mostly on the geometry of the transmission medium, and hence are a slow-time property. They do not need to be estimated as often as intantaneous channel state information, and tracking can be utilized to reduce the complexity of CCM estimation.

Many methods have been proposed to estimate CCMs, and almost all of them are parametric methods. In other words, these methods assume that the CCM pertains to a certain structure, and propose relatively simple methods that are based on this assumption. One such example is [7], where it was assumed that the CCM is a diagonal matrix. In another paper, the sparsity of the CCM was exploited to devise a CFAR-based method for near-optimum CCM estimation [8]. Other authors have utilized the instantaneous CSI to determine the CCM [9].

The estimation of the CCM is a computationally complex task, especially for a system with many antennae. Therefore, CCM estimation is almost never done both in the uplink and the downlink. Some works in the literature propose methods to interpolate the downlink CCM from the uplink CCM, [10], [11], where some works assume TDD operation and perform CCM estimation only in one direction [8].

In this thesis, deep neural networks (DNNs) are utilized to parametrically estimate the CCM of a mmWave MIMO channel. All kinds of machine learning algorithms, including DNNs, have been used to improve performance or decrease complexity in physical layer applications [12].

The term 'machine learning' describes a very diverse set of algorithms that share a common trait: that they are not limited by the commands given to them by their programmers. Supervised learning algorithms, where the computer learns 'patterns' from labeled data, and uses this information to make deductions about new information, is very different from unsupervised learning, where the algorithm decides on the labels based on a user-defined similarity measure. Another major category of machine learning, called reinforcement learning, works on the principle of maximizing 'gains' in a predefined 'game' and is fit to be used in problems much different than those suitable for supervised or unsupervised learning. Due to the vast availability and variety of machine learning tools, they have been applied to many existing problems [13].

Machine learning methods have been used in every mobile communication layer, ranging from physical layer applications to mobile networking problems to app-level applications [14]. Each field of study, although characterized by vastly different requirements and challenges, can employ machine learning and deep learning algorithms to solve a broad range of problems.

For network-level problems, such as routing optimization and scheduling problems, where there is no definitive 'correct' or 'optimum' answer and solutions aim to increase certain performance metrics, reinforcement learning was found to be particularly suitable and found broad popularity [15, 16].

App-level data processing systems, where problems can require predictive qualities or pattern recognition capabilities, were also found to be suitable candidates for machine learning applications. Problems such as network prediction, where long-short-term memory (LSTM) algorithms can be used [17], or traffic classifications problems, where supervised learning algorithms are fit to be used [18], have been found to be suitable applications for machine learning algorithms.

Machine learning and specifically deep learning algorithms have also found popularity in physical layer applications. Many physical layers have been found to be suitable for deep learning applications. For example, the authors of [19] utilize an unsupervised deep auto-encoter in a single-user MIMO system, in order to optimize representations and the encoding/decoding processes.

Deep learning algorihms have also been used in modulation recognition problems, where the modulation of a received signal is unknown and must be determined by the receiver. This is a classification problem that can be solved by supervised learning algorithms, and has been addressed in [20], where convolutional neural networks were used to effectively determine the modulation type of a received signal.

Channel and direction-of-arrival estimation, which are elementary problems in physical layer applications, have been addressed in [21], where DNNs were trained using labeled data obtained through simulations. In this paper, the DNNs were coupled with a proposed tracking method, which aimed to further improve performance. However, due to stringent complexity requirements that accompany instantaneous channel state information estimation, DNNs have not found popular utilization in the estimation of short-term channel characteristics.

DNNs are much more suitable to detect long-term channel characteristics, including the CCM. Estimation problems are particularly suitable for supervised learning problems, since data 'labels', i.e., the true values of the estimated features, are well-defined. Convolutional neural networks have been used in the literature for CCM estimation [22] by assuming correlation between the matrix elements.

In this thesis, on the other hand, instead of trying to find correlation between matrix elements, a parametric estimation of the CCM was done using deep neural networks. Based on the assumption that the CCM fits a specific mathematical model, the deep learning process was greatly simplified, because pattern recognition among matrix elements was no longer necessary.

This thesis is divided into seven chapters. In this chapter, the problem was introduced and a literature review concerning the problem and the method being used, namely deep learning, was presented. In Chapter 2, a more detailed explanation of 5G technologies, along with its benefits and challenges, is given. In Chapter 3, a detailed explanation of machine learning algorithm is given, with an emphasis on deep learning. Types of machine learning algorithms are described and compared, and the training procedure is broadly described. In Chapter 4, The CCM is described in detail. Furthermore, the specific mathematical model that is used in this thesis is provided and a justification of the selected model is given. In Chapter 5, the main contribution of this thesis is presented. The system model being used and the methods that are proposed for CCM estimation are described in detail. The performances of the proposed methods are discussed in Chapter 6, with simulation results and comments. Chapter 7 concludes this thesis, with a brief overview of this thesis, the proposed methods, and the results.

# CHAPTER 2

# 5G – NEW RADIO

In this chapter, 5G technologies are described. A brief summary of mobile communication standards (generations) is given, and 5G is compared with previous generations of communications standards. Three main categories of 5G use cases is described and the chapter is concluded by discussing two key enabling technologies of 5G.

## 2.1  A Brief History of Mobile Communications

The importance and the everyday use of wireless communications has been increasing since the turn of the twentieth century. Wireless communications have helped shape modern history since the invention of the radio, and continue to do so, with an ever-increasing impact. The utilization of wireless communications in day-to-day life has become even more widespread since the introduction of cellular phones in the 1980s, whose popularity contributed to accelerating the development of communication technologies even further.

In 1991, second-generation (2G) cellular technology was introduced in Finland on the Global System for Mobile Communications (GSM) standard. Ever since the introduction of the GSM standard, a new generation of wireless communication standards has evolved roughly every ten years, each generation bringing with it a unique aspect.

In first generation mobile communications, the voice signal was transferred in an analog format. In 2G systems (GSM), voice data transfer became digital. With the introduction of Enhanced Data Rate for Global Evolution (EDGE) technology, also called 2.5G, data rates for digital communication were increased.

In the beginning of the 2000s, 3G standards were introduced, with Wireless Code Division Multiple Access (WCDMA) as a key enabling technology. 3G allowed mobile phones to connect to and use the internet.

However, it was not until 4G (also called Long Term Evolution – LTE) was introduced that mobile internet became a primary aspect of mobile phone use. 4G brought with it increased data rates which, along with the introduction of the smartphone, made internet connection the primary use of mobile phones. 4.5G (LTE-Advanced, LTE-A) technologies, while not bringing with them major paradigm shifts compared to 4G technologies, drastically increased data rates and therefore contributed to the widespread use of mobile internet.

Similar to all of its predecessors, 5G – New Radio comes as a response to much different needs an therefore comes with its own unique features, which are detailed in the next section. An overview of all modern digital communication standards can be seen in Figure 2.1.

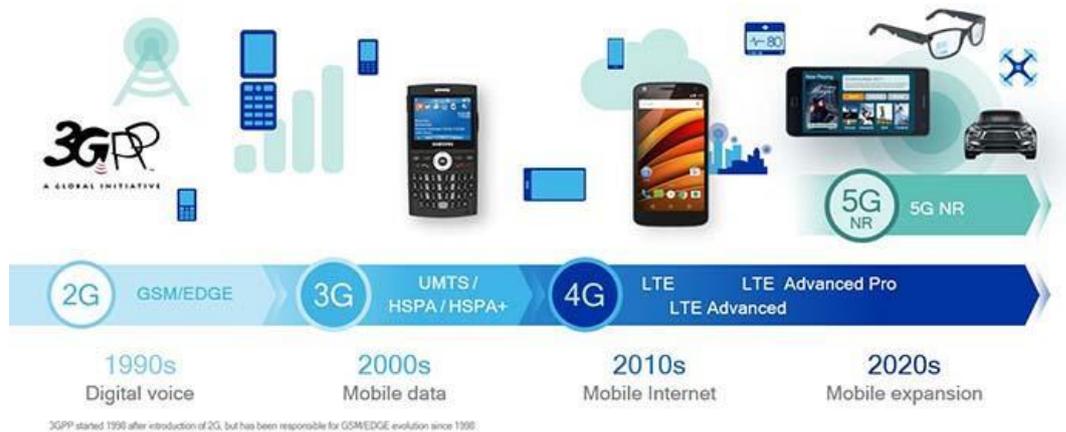

Figure 2.1 An overview of all modern digital communication standards, from 2G to 5G [23].

## 2.2    5G and Its Primary Features

In the past years, the increasing number of digital connectivity of everyday items (such as kitchenware, smart watches, sensors, etc.) and the emergence and growth of the Internet of Things (IoT) have brought with them new challenges and requirements for digital communications.

Similar to previous generations, 5G will bring with it dramatically increased data rates. For mobile phones, internet speeds are expected to increase by at least one order of magnitude. However, that is not the only noteworthy feature of 5G.

Unlike previous generations, 5G is expected to support a very heterogeneous network of devices, where each user will have vastly different requirements. For example, a mobile phone that is streaming a 4K resolution video will be connected to the same network as a self-driving car, which may not need the same connection speed as the mobile user but needs important messages to be delivered immediately. The 5G network will have to simultaneously meet the needs of all these different users, which means that 5G networks need to operate in a highly dynamic fashion, adapting to each user's requirements as needed.

To account for this heterogeneity, three main categories of use cases for 5G were identified. These categories are defined in the section below.

## 2.3  5G Use Cases

Three main categories of use cases have been defined for 5G, each with its unique requirements and challenges. These use cases are Enhanced Mobile Broadband (eMBB), Ultra Reliable Low Latency Communication (URLLC) and Massive Machine Type Communication (mMTC) [24].

### 2.3.1  Enhanced Mobile Broadband (eMBB)

Enhanced Mobile Broadband is the category representing day-to-day mobile internet use, primarily through mobile phones and personal computers. eMBB is characterized by:

- Extremely high data rate requirements,

- Comparatively soft latency requirements,
- Comparatively soft reliability requirements and
- A medium amount of users.

### 2.3.2  Ultra Reliable Low Latency Communication (URLLC)

Ultra Reliable Low Latency Communication is the category representing high-priority and high-sensitivity connection needs, such as the remote operation of machines (e.g., remote medical operations) and intelligent transportation (e.g., self-driving cars and smart trains). URLLC is characterized by:

- Medium data rate requirements,
- Very strict latency and reliability requirements and
- A relatively low amount of users.

### 2.3.3  Massive Machine Type Communication (mMTC)

Massive Machine Type Communication is designed to serve a very large amount of users that do not require high data rates. mMTC accounts for the mobile connection of IoT devices, such as sensors in a factory or household objects. mMTC is characterized by:

- Very low data rate requirements,
- Relatively soft latency and reliability requirements and
- An extremely high amount of users.

These three use cases are generally represented with the famous "5G pyramid", given in Figure 2.2, which helps to visually compare their features and sample scenarios where they will be used.

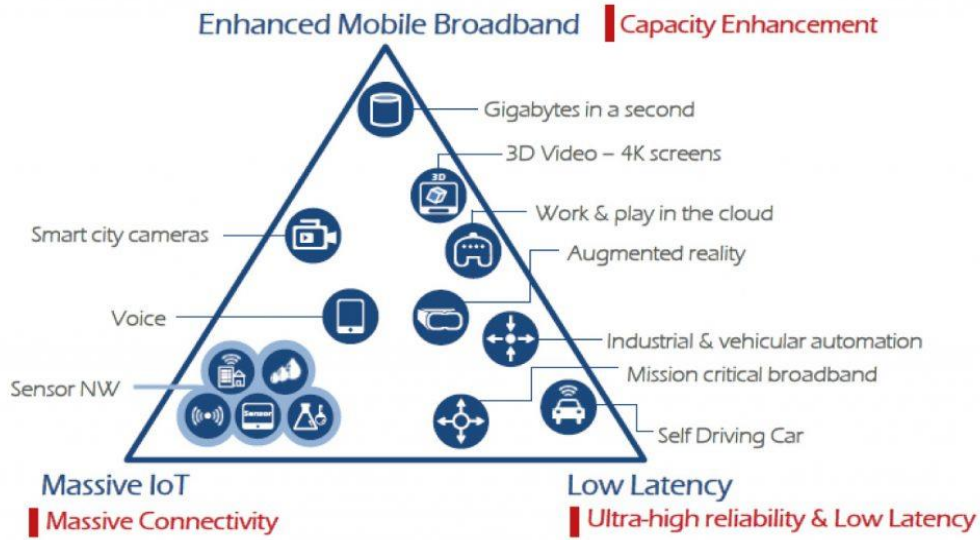

Figure 2.2 The 5G pyramid [25].

**2.4  5G Enabling Technologies**

5G technologies rely heavily on existing technologies, mainly those of 4G networks, such as orthogonal frequency division multiple access (OFDMA), sub-6GHz communications, and cellular networks. However, since 5G is expected to fulfill new and more demanding requirements, it also needs new technologies that were not present in previous generations. Similar to all of its predecessors, 5G comes with some new enabling technologies. The main enabling technologies for 5G are massive MIMO and millimeter-wave communications.

**2.4.1  Massive MIMO**

Without spatial diversity, the capacity of a wireless communication scheme is limited by diversity in time and frequency domains. In fact, the capacity of a wireless channel can be roughly determined by the multiplication of its bandwidth and time duration [2],

$$C \cong WT.$$

Since time and frequency are limited resources for a given communications scenario, these bounds can be considered as hard limits on data rates. As can be expected, modern communications schemes (up to 4G) have been operating very close to this limit. Therefore, to meet the demand for increased data rates, communications engineers have turned to utilize diversity in a new dimension, namely in the spatial domain. A wireless communications system can benefit from spatial diversity with the help of multiple antennas, which can be placed at the transmitter or the receiver side (or both).

Multiple antennas can be used in multiple ways to achieve increased $SE$:

- Analog beamforming can be used to steer the antenna beams towards specific users, which reduces the correlation between the channels of different users [26]. Analog beamforming is done by applying phase shifts to antenna arrays and is an existing method that has been used by antenna engineers since the beginning of the twentieth century. A representation of analog beamforming can be seen in Figure 2.3.

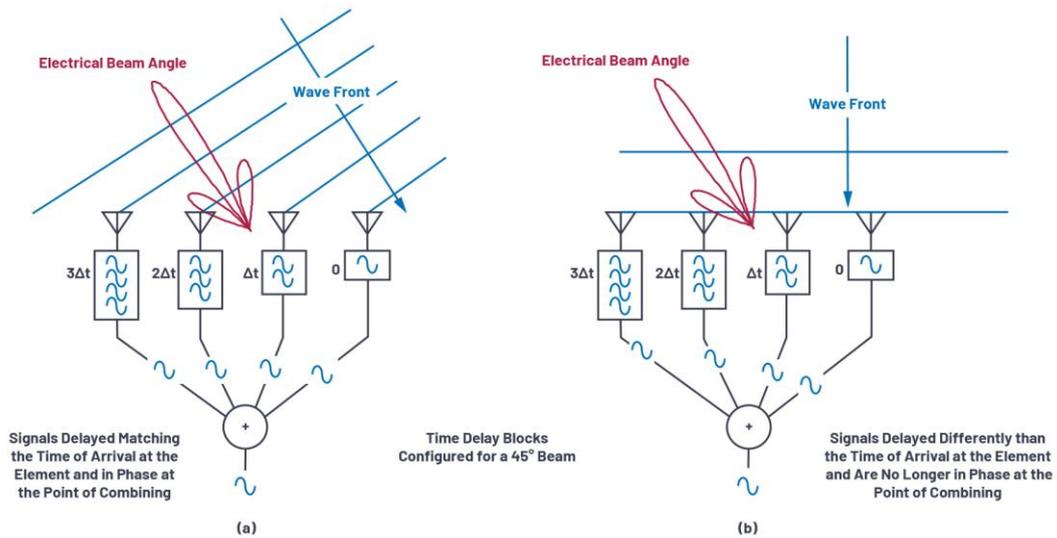

Figure 2.3 Analog beamforming/beam steering with a phased antenna array [26].

- Digital beamforming can be used to exploit instantaneous channel realizations to isolate the channels of each user. Digital beamforming

requires the use of many RF chains and is therefore more costly than analog beamforming. However, digital beamforming can isolate user channels much more precisely when the channel is accurately known.

- Spatial multiplexing, which is the transmission of multiple data streams from multiple antennas or antenna arrays in a parallel manner, can be used to increase data rates.

Multi-input-multi-output (MIMO) communications have been studied since the 1970s [27] and they were even included in the LTE-A standard [28], but their use has been limited before 5G.

In 4.5G communications, 4x4 MIMO (constituting four transmitting antennas and four receiving antennas) was included in the standard. This enabled 4.5G networks to benefit from spatial diversity to improve error performance and to employ spatial multiplexing to increase data rates. However, the relatively low number of antennas meant that the benefits of MIMO were also limited.

Compared to 4G and 4.5G networks, 5G MIMO systems will use a much larger number of antennas, also called massive MIMO (mMIMO) systems. Common 5G configurations include as many as 192 antennas and up to 64 RF chains at the BS side.

Using this many antennas brings some challenges, some of which have been solved and some of which have been only partially mitigated. Some of the challenges of mMIMO systems are given below.

- Since each transmitting-receiving antenna pair has its own channel, it is challenging to perform channel estimation in mMIMO systems. A partial solution for this challenge was to use time division duplex (TDD) systems, where the uplink and downlink channels can be assumed to be the same, which enables channel estimation to be performed only on one direction and used for both directions, thereby reducing the computational load and training overhead by half.

- Using a large number of RF chains is a computationally heavy task, but this problem is alleviated by the increased computational capabilities of modern chips. Using analog and digital beamforming together (called hybrid beamforming) is another way of finding a good trade-off between good performance and low complexity.
- Large antenna arrays are not always easy to cool down. This is a problem for the antenna engineers and techicians who design the BS hardware to solve.

Acceptable solutions for all challenges regarding mMIMO were found, enabling 5G BSs to employ mMIMO as a key enabling technology.

### 2.4.2 Millimeter Wave Frequencies

In mobile communication standards up to 5G, Sub-6GHz bands are used. This is because of the favorable propagation properties of these bands and because the bandwidth available in these bands has been sufficient in satisfying network bandwidth requirements. However, with increasing demand for bandwidth, these bands are no longer sufficient.

As stated in the previous section, the capacity of a wireless channel is directly proportional to the bandwidth occupied by that channel. It is also not hard to see that lower frequency bands have lower bandwidths available for communication. (For example, it is not a huge challenge to allocate a 500MHz bandwith in the 50 GHz band, whereas 500MHz is an extremely large bandwith in the 3GHz band.) That is why 5G communications employ millimeter-wave (mmWave) bands (frequency bands where wavelengths are on the order of millimeters) for high speed wireless communications. Some mmWave frequency bands specified in 5G standards are 26.5–29.5 GHz, 24.25–27.5 GHz, 39.5–43.5 GHz, 37–40 GHz and 27.5–28.35 GHz [29].

mmWave bands have large amounts of unutilized bandwidths, which directly makes much higher data rates achievable. However, mmWave also has its drawbacks. Some of these drawbacks are given below [1].

- Electromagnetic wave attenuation is directly proportional to operation frequency. Antenna signals at mmWave frequencies attenuate much more quickly than those at Sub-6GHz frequencies, which is problematic for coverage concerns. In 5G, small cells will be used in mmWave bands. Small cells will be deployed in areas where there is a high density of users and a high density of BSs, such as central urban areas. Sub-6GHz bands will continue to be used where a low density of BSs is sufficient to serve a large area, such as rural areas.
- Electromagnetic waves are more prone to absorption in mmWave bands. This also prevents mmWave bands to be used in scenarios where large coverage areas are required. This challenged can also be overcome with the methods described above.
- The size of an antenna has to be proportional with the wavelength of the transmitted wave. Therefore, mmWave antennas are about an order of magnitude smaller than Sub-6GHz antennas, which is problematic in terms of energy transmission. Smaller antennas cannot radiate as much power as their larger counterparts, which presents a further problem in terms of cell radius requirements. This challenge was overcome with the use of large-scale antenna arrays, i.e., MIMO communications, which not only increase the amount of energy radiated from the antennas, but also increase the directivity of the radiated beams, which boosts the coverage even further.

As mentioned before, mmWave bands will not be the only frequency bands used in 5G networks. Sub-6GHz bands will continue to exist and supplement mmWave bands where mmWave bands are not suitable for the given use case.

# CHAPTER 3

# MACHINE LEARNING AND DEEP NEURAL NETWORKS

In this chapter, machine learning algorithms are discussed with and emphasis on deep learning algorithms in particular. Three major categories of machine learning algorithms are described. Deep learning algorithms are defined and their general characteristics are discussed. Finally, an explanation of the deep neural network training procedure is given.

## 3.1 Machine Learning

Machine learning has been in the spotlight of the scientific community for the past few years. Despite its recent rise in popularity, machine learning has been around for more than half a century. The term 'machine learning' was coined in 1959 [30], and there has been continuous interest in the field since its conception. However, due to a lack of high-performance computers before the twenty-first century, studies on machine learning remained largely theoretical and no large-scale implementations were possible until the 1990s.

In recent years, due to the abundance and low cost of computational power, machine learning algorithms have once more come to popularity, finding areas of use in practically every scientific field.

The terms 'machine learning' and 'artificial intelligence' have been used in a very loose sense, and it may not always be clear what exactly constitutes machine learning. Generally, the term 'machine learning' is used to describe any kind of algorithm where the action is not deterministically dictated by the programmer. Most machine learning algorithms are closely related to optimization and pattern recognition algorithms, which make them familiar to an electrical engineer who has worked in any one of the fields optimization, telecommunications, signal processing or pattern recognition.

Machine learning algorithms are usually divided into two broad categories: supervised learning, unsupervised learning and reinforcement learning.

### 3.1.1 Supervised Learning

Supervised learning describes a class of machine learning algorithms where the data is definitively labeled. In other words, the machine learning algorithm receives a set of 'correct answers' with the training data, which it uses to learn from. The 'cost' and 'reward' for supervised learning algorithms are measures of how closely the algorithm can match these correct answers.

Supervised learning algorithms make up the most popular algorithms that are used in scientific literature, mainly because they are simple to implement and straightforward to understand. In many scientific contexts, labeled data is either readily available or easy to produce, making supervised learning an attractive choice for scientists.

Supervised learning algorithms naturally also have their drawbacks. For example, in scenarios where it is not always possible to produce labeled data due to physical constraints, supervised learning will not be a suitable choice. Another limitation is that the performance of supervised learning algorithms rely very heavily on the correctness and the quality of the labeling. If there are any errors in the labeling of the data, a supervised learning algorithm cannot account for these errors, as it blindly trusts its training data.

Applications of supervised learning algorithms can be divided into two broad categories, namely classification problems and regression problems. Classification problems are problems where the algorithm makes a decision among a finite number of choices according to the input. An example can be an algorithm that decides whether a given image belongs to a cat, a dog or a bird. Since the algorithm decides between three choices, this is a classification problem. Regression problems are problems where the algorithm creates a model of the problem that fits the training data and applies this model to new inputs. An example can be predicting the price of a house given its distance to the center, number of rooms, and the age of the building. Since the price of a house is a function of these parameters and is not selected from a finite set of values, this is a regression problem. It can be seen that regression problems are similar to function fitting problems.

Algorithms that can be used for supervised learning include linear regression, logistic regression, K-nearest neighbor algorithm, artificial neural networks and convolutional neural newtorks [31].

### 3.1.2 Unsupervised Learning

Unsupervised learning describes a class of machine learning algorithms that learn patterns from unlabeled data. Some applications of unsupervised learning are

- Clustering, where the algorithm divides the input data into groups while trying to maximize the similarity of data samples within groups. The algorithm determines the grouping rule according to the similarity measures defined by the user. Examples for clustering methods include k-means clustering and hierarchical clustering [32].
- Anomaly detection, where the algorithm aims to find data samples that are outside of the norm of the whole dataset. The algorithm learns what constitutes as 'normal' from the samples in the training data. Anomaly detection can be done through various methods, such as support vector machines, rule-based anomaly detection and statistical anomaly detection [33].

### 3.1.3 Reinforcement Learning

Reinforcement learning describes a class of machine learning algorithms where instead of being trained on a dataset, the algorithm is given a set of rules and develops methods to maximize a predetermined 'reward function'. Reinforcement learning is more suitable for problems that include a short-term vs long-term reward trade-off.

Some examples where reinforcement learning was used to obtain satisfactory results include classic boardgames such as checkers, backgammon, and chess.

Reinforcement learning can be implemented with various approaches, such as dynamic programming, temporal-difference learning and Monte Carlo methods [34].

## 3.2 Deep Learning

In this section, Deep Learning algorithms and deep neural networks are discussed in detail. Deep neural networks are a subclass of a machine learning method called artificial neural networks. For simplicity, single-layer artificial neural networks are described before going into more detail about deep neural networks.

### 3.2.1 Artificial Neural Networks

Artificial neural networks (ANN) are a popular class of supervised learning algorithms, where the output is a function of a linear combination of some input variables. ANNs can be used in both classification and regression problems. A sample ANN can be seen in Figure 3.1, where each input node represents an independent variable that affects the output, and each output node represents a prediction/decision about a dependent variable.

The reason ANNs are named after neurons is because the basic idea behind ANNs is inspired by neurons. In a neurological system, each neuron (each cell) is connected to other neurons (cells) through synapses, and the number and strength of these connections determines the behavior of the neurogical sytem. Similarly, in an ANN, nodes are connected to other nodes and these connections determine the operation of the neural network.

A mathematical description of an ANN with $n_{in}$ input nodes, $n_{hidden}$ hidden nodes and $n_{out}$ output nodes is

$$\boldsymbol{w}_{out} = \begin{bmatrix} w_1^{(out)} \\ w_2^{(out)} \\ \vdots \\ w_{n_{out}}^{(out)} \end{bmatrix} = f_2(\boldsymbol{W}^{(2)}\boldsymbol{a})$$

$$= f_2\left(\begin{bmatrix} w_{11}^{(2)} & w_{21}^{(2)} & \cdots & w_{1n_{hidden}}^{(2)} \\ w_{12}^{(2)} & w_{22}^{(2)} & \cdots & w_{2n_{hidden}}^{(2)} \\ \vdots & \vdots & \ddots & \vdots \\ w_{n_{out}1}^{(2)} & w_{n_{out}2}^{(2)} & \cdots & w_{n_{out}n_{hidden}}^{(2)} \end{bmatrix} \begin{bmatrix} a_1 \\ a_2 \\ \vdots \\ a_{n_{hidden}} \end{bmatrix}\right),$$

$$\boldsymbol{a} = \begin{bmatrix} a_1 \\ a_2 \\ \vdots \\ a_{n_{hidden}} \end{bmatrix} = f_1(\boldsymbol{W}^{(1)}\boldsymbol{w}_{in})$$

$$= f_1\left(\begin{bmatrix} w_{11}^{(1)} & w_{21}^{(1)} & \cdots & w_{1n_{in}}^{(1)} \\ w_{12}^{(1)} & w_{22}^{(1)} & \cdots & w_{2n_{in}}^{(1)} \\ \vdots & \vdots & \ddots & \vdots \\ w_{n_{hidden}1}^{(1)} & w_{n_{hidden}2}^{(1)} & \cdots & w_{n_{hidden}n_{in}}^{(1)} \end{bmatrix} \begin{bmatrix} w_1^{(in)} \\ w_2^{(in)} \\ \vdots \\ w_{n_{in}}^{(in)} \end{bmatrix}\right),$$

where the input to the algorithm is the $n_{in} \times 1$ vector $\boldsymbol{w}_{in}$ containing the independent variables and the output of the algorithm is the $n_{out} \times 1$ vector $\boldsymbol{w}_{out}$ containing the dependent variables.

Assuming that the outputs are a function of the inputs, the machine learning algorithm essentially learns the system model, i.e., learns the function that relates the inputs and the outputs, by determining what the weight matrices $\boldsymbol{W}^{(1)}$ and $\boldsymbol{W}^{(2)}$ should be.

The purpose of the functions $f_1$ and $f_2$ is to introduce nonlinearity to the model. These functions are called 'activation functions'. The choice of activation functions is a design parameter. Popular choices for $f_i$ are the sigmoid function, $S(x)$, and the rectifier function, $R(x)$, whose definitions are

$$S(x) = \frac{1}{1 + e^{-x}}, \qquad R(x) = x^+ = \max(0, x).$$

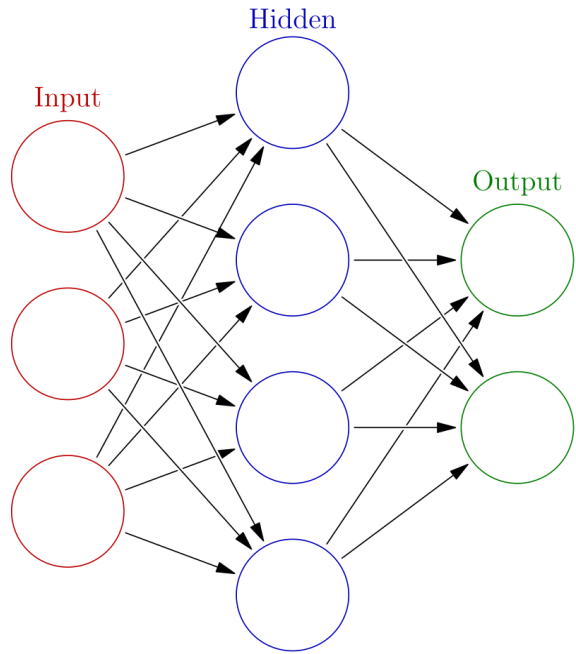

Figure 3.1 A sample artificial neural network, with three input variables, four hidden nodes and two outputs.

The training process, i.e., the process where the machine learning algorithm determines the values of the elements of $\boldsymbol{W}^{(1)}$ and $\boldsymbol{W}^{(2)}$, is described in the following sections.

There exist many types of ANNs, each of which have evolved from the basic single-layer ANN to suit different needs and to perform different tasks. Two of these are deep neural networks and convolutional neural networks (CNN). Both algorithms are described in the following sections.

Deep neural networks were utilized as a main part of the method developed for this thesis, but CNNs were found to be unsuitable for the problem at hand.

**3.2.2  Deep Neural Networks and Deep Learning**

Deep learning is the name given to a subclass of supervised learning algorithms that employ artificial neural networks with multiple hidden layers. In other words, deep neural networks (DNNs) are a generalization of single-layer ANNs. A sample neural network can be seen in Figure 3.2.

The input-output relation for a DNN with $N_L$ hidden layers is

$$\boldsymbol{w}_{out} = \boldsymbol{a}^{(N_L+2)} = f_{N_L+1}\big(\boldsymbol{W}^{(N_L+1)}\boldsymbol{a}^{(N_L+1)}\big),$$

$$\boldsymbol{a}^{(N_L+1)} = f_{N_L}\big(\boldsymbol{W}^{(N_L)}\boldsymbol{a}^{(N_L)}\big),$$

$$\ldots$$

$$\boldsymbol{a}^{(2)} = f_2\big(\boldsymbol{W}^{(1)}\boldsymbol{w}_{in}\big).$$

It can be easily seen that the mathematical expression for DNNs is very similar to that of ANNs, only with a higher number of operations.

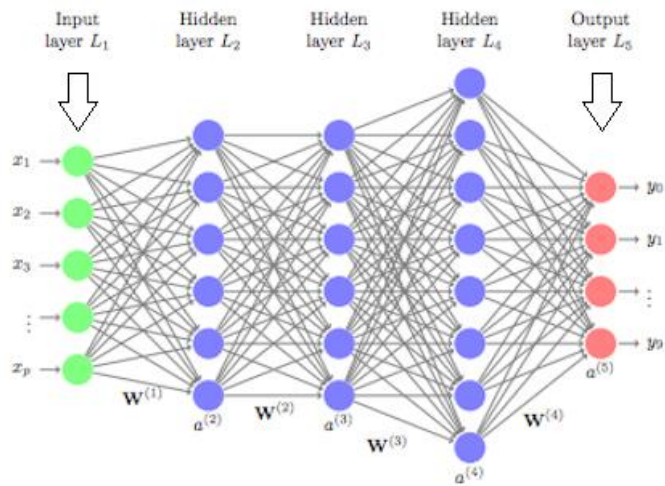

Figure 3.2 A sample neural network layout, with three hidden layers.

Even though DNNs are mathematically very similar to single-layer ANNs, the two classes of machine learning algorithms have significant differences. Some of these differences are explained below.

- Single-layer ANNs are very senstitive to the selection of input variables. Returning to one of the above examples, consider a single-layer ANN that is being used to predict the prices of houses. Whether the area of the houses is represented in square-meters ($m^2$) or in square-feet (sqft) will dramatically affect the performance of the ANN. On the other hand, DNNs are much more robust (but far from immune) to these kinds of varieties. This is because after training, some of the hidden layers can act as normalizers and offset this effect. However, this is not controllable or guaranteeable, therefore the selection of input variables is also an important problem for DNNs [35].

- Single-layer ANNs can only model 'simpler' functions. Since the mathematical representation of ANNs is basically two linear transformations coupled with two nonlinear transformations, they can only model systems that have a form that either fits or resembles this structure. DNNs share the same limitation, but since they have multiple hidden layers, DNNs can account for nonlinearities much more easily, and can therefore represent a much broader group of systems.
- Since single-layer ANNs have a simpler structure, fewer data samples are enough to train them. Since DNNs are much more complicated, they are much more prone to overfitting [35] if a large number of samples is not available to them for training. The concept of overfitting is discussed in the following section.

### 3.2.3 Neural Network Training

As stated before, a deep learning algorithm aims to determine the elements of the weights matrices $\boldsymbol{W}^{(1)}, \dots, \boldsymbol{W}^{(N_L+1)}$ so that the DNN can accurately represent the function relating the given inputs and the outputs.

The first step in determining these weights is to define a performance metric. Without a performance metric, the algorithm has no sense of what a succesful result is. The performance metric provides the machine learning algorithm with a concrete/numerical measure of how succesful it is.

The most common performance metric used in machine learning algorithms is the mean square error (MSE). MSE is a widely used performance metric in many fields of electrical engineering and is defined as

$$MSE = \frac{1}{n}\sum_{i=1}^{n} |\boldsymbol{y}_i - \widetilde{\boldsymbol{y}_i}|^2$$

$$= \frac{1}{n}\sum_{i=1}^{n}(y_i - net(x_i))^2,$$

where $n$ is the number of data samples (i.e., the size of the training set) $y_i$ is the vector containing the DNN outputs for the $i$th data sample, $x_i$ is the vector DNN inputs for the $i$th data sample, and $net(.)$ represents the operation performed by the neural network.

Since the training samples $\{x_i, y_i\}_{i=1}^{n}$ are fixed, $MSE$ can be considered as a function of the weight matrices $W^{(1)}, \ldots, W^{(N_L+1)}$. Since the function $net(.)$ is an explicit function, the cost function is also an explicit function of the DNN weights. However, this function is a very complicated function that has extremely many dimensions. In fact, if each matrix $W^{(i)}$ has dimensions $N^{(i)} \times M^{(i)}$, then the cost function is a function over a space of $\sum_{i=1}^{N_L} N^{(i)} M^{(i)}$ dimensions. The deep learning algorithm tries to find the minimum value of the cost function over this complicated space.

### 3.2.3.1 The Gradient Descent Method

Minimizing a well-defined, explicit function over a given space is a problem very familiar to many electrical engineers, especially those who work in the fields of signal processing and optimization. A very well-known optimization method, gradient descent, is also used in the field of machine learning to determine the optimum weights of the weight matrices [36]

To iteratively find the point minizing a multi-variate function $F(x)$, the following iterative algorithm can be used:

$$a_{n+1} = a_n - \gamma \nabla F(a_n),$$

where $\gamma \in \mathbb{R}$ is a design parameter called the 'scaling factor' (or the 'step size' or the 'learning rate'), and $\nabla$ is the gradient operator, which represents the direction where the slope of the function $F(x)$ is the highest. Going on the exact opposite direction means moving in the direction with the lowest slope, i.e., the direction that most quickly decreases $F(a_n)$.

For a more intuitive reasoning, consider a cost function over two dimensions, a representation of which is given in Figure 3.3. Moving in the direction with the lowest slope, the point that minimizes the given function is detected in a few steps.

the gradient descent method is guaranteed to find the point $a_0$ minimizing any convex function[2] $F(x)$ given that $\gamma$ is small enough.

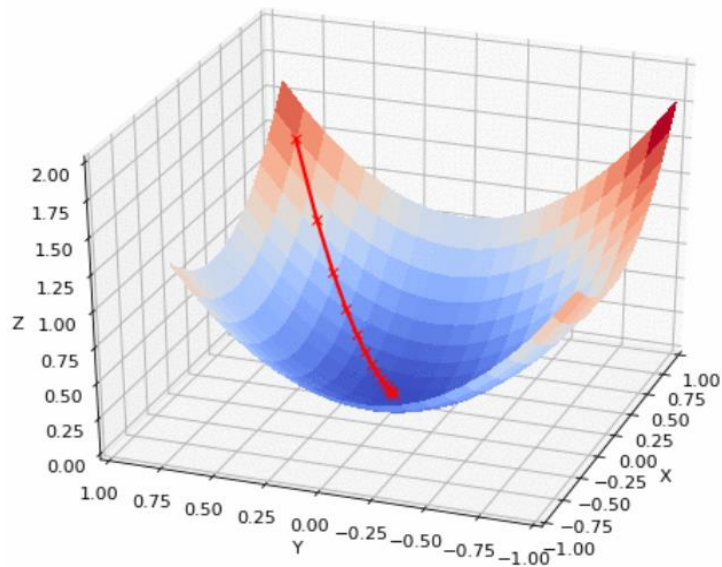

Figure 3.3 An example of the gradient descent algorithm [37].

However, the gradient descent method cannot be guaranteed to find the point $a_0$ that minimizes $F(x)$ if $\gamma$ is too large or if $F(x)$ is not a convex function, i.e., if it has multiple local minima [36].

A visualization of the cases where $\gamma$ is too large is given in Figure 3.4. It can be seen in the figure that the large step sizes cause the algorithm to keep 'stepping over' the desired minimum point and therefore the points $a_n$ cannot converge.

---

[2] A convex function is a function $f(x): X \rightarrow \mathbb{R}$ that satisifes the inequality
$$f(tx_1 + (1-t)x_2) \leq tf(x_1) + (1-t)f(x_2)$$
for all $t \in [0,1]$ and all $x_1, x_2 \in X$.

The other case where gradient descent cannot find the global minimum is visualized in Figure 3.5. Here, gradient descent cannot be guaranteed to find the global minimum, because it converges to one of the local minima. Since the algorithm moves towards the direction that has the lowest slope, it cannot 'escape' local minima once it converges.

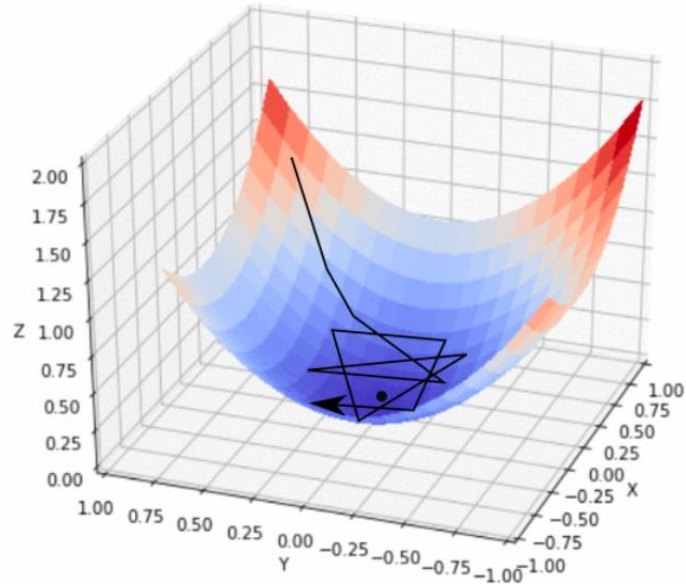

Figure 3.4 An example of the gradient descent algorithm where $\gamma$ is too large [37].

This problem is frequently experienced in deep learning training, since the cost functions in deep learning are typically not convex. Some of the techniques typically used in overcoming this problem are:

- Performing multiple trainings, each with a different (randomly selected) starting point, and selecting the network that yields the best result. This approach helps the algorithm find multiple local minima since the trajectory of the algorithm will be different for each starting point.
- Using gradient descent with momentum, where a fraction $\eta \boldsymbol{a}_n$ of the previous update vector $\boldsymbol{a}_n$ is added to the new update vector $\boldsymbol{a}_{n+1}$ at each iteration. The 'momentum', similar to physical momentum, helps the algorithm cross small 'hills' at the cost of increased convergence time. The mathematical expression of this method is as follows:

$$Step\ 1: \boldsymbol{a}_n = \eta \boldsymbol{a}_{n-1} + \gamma \nabla F(\boldsymbol{\theta}),$$

$$Step\ 2: \boldsymbol{\theta} = \boldsymbol{\theta} - \boldsymbol{a}_n.$$

There are many more variants of the steepest descent method in the literature, each of which was proposed to overcome a different problem [38]. However, since they are too many to list, only the two methods that were used in this work were described.

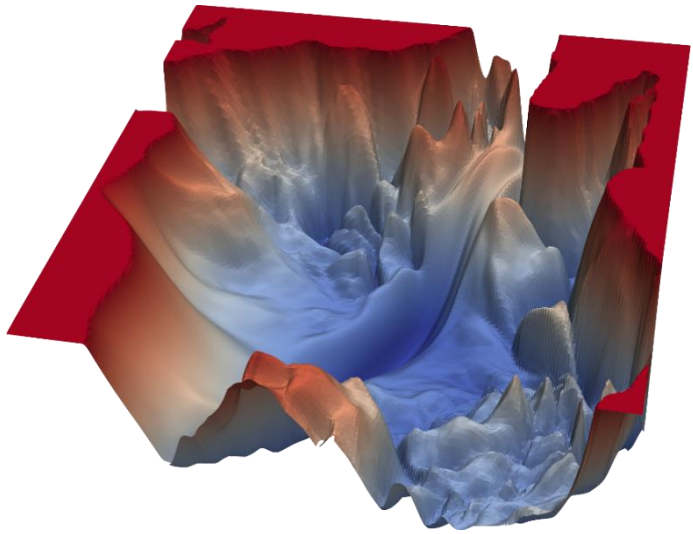

Figure 3.5 A sample cost function where there are multiple local minima, because the cost function is not convex. [37]

### 3.2.3.2 Backpropagation Algorithm

One of the main restrictive factors concerning machine learning (and one of the main reasons it has not gained popularity until decades after its introduction) is computational complexity. Even though the update terms for the gradient descent algorithm are well-defined, they are very expensive to compute because of the sheer number of multiplications involved in the matrix multiplication operation.

To somewhat alleviate this problem, the backpropagation algorithm was proposed in the 1980s in the seminal paper [39]. The backpropagation algorithm is a method for efficiently calculating the update terms for the gradient descent algorithm. Its main feature is its computational efficiency, and it does not yield different results compared to other methods that determine the update terms for gradient descent.

The main idea behind this algorithm is to start from the error at the output layer and to 'propagate' the error through each layer according to the 'error' on each node on each layer and the derivatives of the activation functions $f_i$.

The backpropagation algorithm is immensely popular and is used in virtually every deep learning environment/library that exists. Since the mathematical/algorithmic expression is very complicated and also not relevant to this paper, it was not presented here.

# CHAPTER 4

## CHANNEL COVARIANCE MATRIX ESTIMATION

In this chapter, channel covariance matrices (CCMs) are described. Some properties and uses of CCMs are given and a literature survey about the problem of channel covariance matrix estimation is presented. Finally, a mathematical justification of the CCM model used in this thesis is provided.

### 4.1 Channel Covariance Matrices

In a MIMO channel, the channel covariance matrix (CCM) is the matrix containing the covariance of each pair of channel instances of the channel. In other words, the CCM contains statistical information about the channel, which can be utilized to improve the connection quality.

Consider an $N \times K$ MU-MIMO channel with $K$ single-antenna users and one $N$-antenna base station. For a single user, the uplink channel can be represented by an $N \times 1$ column vector $\boldsymbol{h} = [h_1 \quad h_2 \quad ... \quad h_N]^T$, where $h_i$ are random variables with means $\{\mu_i\}_{i=1}^{N}$. Then, the CCM of this user is defined as

$$\boldsymbol{R} \triangleq E\{(\boldsymbol{h} - \boldsymbol{\mu})(\boldsymbol{h} - \boldsymbol{\mu})^H\}, \tag{4.1}$$

where $\boldsymbol{\mu} = E\{\boldsymbol{h}\} = [\mu_1 \quad \mu_2 \quad ... \quad \mu_N]^T$.

To clarify, $\boldsymbol{R}$ is an $N \times N$ matrix whose $ij$th element is the covariance of the $i$th element and the $j$th element of $\boldsymbol{h}$,

$$R_{ij} = Cov(h_i, h_j) = E\{(h_i - \mu_i)(h_j - \mu_j)^*\}. \tag{4.2}$$

In practical cases (and in this thesis), the channels are considered to be zero-mean variables, i.e., $\mu_i = 0, \forall i \in \{1, ..., N\}$. Then, Eqns. (4.1 and 4.2 can be rewritten as

$$\boldsymbol{R} \triangleq E\{\boldsymbol{h}\boldsymbol{h}^H\},$$

and

$$R_{ij} = Cov(h_i, h_j) = E\{h_i h_j^*\}.$$

For a spatially white zero-mean channel, since the covariance of any pair of channels is equal to zero, the CCM is a diagonal matrix whose diagonal elements represent the power of each channel. In scenarios where the receiving antennas experience correlated channels, the CCM is not a diagonal matrix. In fact, for highly correlated channels, such as mmWave channels, the CCMs are usually low rank or sparse matrices. This property can be exploited to reduce the dimension of the received signal and reduce computational costs at the BS.

The CCM has other uses that make it an important slow-time property. For example, the CCM can be used in improving the channel estimation performance for spatially non-white channels. Statistical information about the channel can also be used to improve the performances of other operations in the BS, such as beamforming and SNR prediction [6]. There are many works [4], [5] in the literature that utilize the knowledge of the channel statistics, i.e., the CCM, to overcome the challenges of massive MIMO, such as prohibitive computational complexity, uplink pilot contamination, and training overhead.

### 4.2 CCM Estimation

A common problem in massive MIMO communications, computational complexity, also affects the CCM estimation problem. For an $N \times K$ multi-user massive MIMO channel, the CCM estimation accounts for the estimation of two $N \times N$ matrices (one for the uplink, one for the downlink) for each of $K$ users. It is an even bigger challenge to estimate downlink CCMs due to the finite feedback capacities of the users.

This limitation is typically overcome by using time division duplex (TDD) operation. In TDD operation, the MIMO channel is practically reciprocal, i.e., the uplink and downlink channels are very close or the same, which means that uplink CCM estimation is sufficient for use in both the uplink and the downlink. However, methods for frequency division duplex (FDD) operation have also been proposed, such as a method for interpolating the downlink CCM from the uplink CCM [10], [11].

Another way to reduce the complexity of the CCM estimation problem is to use parametric methods. Parametric methods are a class of estimation methods where an assumption is made about the structure of the estimated property and the estimation is done based on this assumption. For example, in [7], the CCM estimation problem was greatly simplified by assuming that the CCM is a diagonal matrix. The authors claim that this is a reasonable assumption for certain array geometries where the diagonalization of the CCM is possible. However, one can note that this assumption somewhat limits the number of scenarios where this method is applicable.

In this thesis, the CCM is assumed to belong to a sparse mmWave channel and is assumed to be a function of the angles of arrival (AoA) and the angular spreads of the user beams. This CCM model, described in detail in Chapter 5, was used in several works in the literature. One such work is [9], where the angle parameters and the power angular spectrum of the channel are determined by using the instantaneous channel state information (CSI).

## 4.3  The Mathematical Model of the CCM

In this thesis, the estimation of the CCM was performed parametrically. In particular, the following model for CCMs, which can be found in [6], was used. Consider an uplink transmission system with $K$ users, where each user $k$ has $L_k$ delay taps. For the $l$th delay tap of the $k$th user, let the AoA be $\theta_k^l$, the angular spread be $\sigma_k^l$, and the beam power be $\rho_k^l$. Then, the CCM of the $l$th delay tap of the $k$th user, $\boldsymbol{R}_k^l$, is given by the equation

$$\boldsymbol{R}_k^l = \rho_k^l \int_{\theta_k^l - \frac{\sigma_k^l}{2}}^{\theta_k^l + \frac{\sigma_k^l}{2}} \boldsymbol{u}(\phi)\boldsymbol{u}^H(\phi) d\phi, \qquad 4.3$$

where

$$\boldsymbol{u}(\phi) = \frac{1}{\sqrt{N}} \left[1, e^{j\pi \sin(\phi)}, \ldots, e^{j\pi(N-1)\sin(\phi)}\right]^T$$

is the unit-energy steering vector for a ULA with an antenna spacing of $d = \lambda/2$. A justification of this CCM model is given below.

### 4.3.1  Channel Covariances for a Single Beam

For a more intuitive explanation of this model, consider the following scenario. Let an $N \times 1$ ULA array stand in front of a plane wave, at an angle of $\theta$ with the wavefront, depicted in Figure 4.1. The wavefront travels a distance of $d(n-1)\sin\theta$ from the first antenna to the $n$th antenna. Then, taking the first antenna in the array as a reference point (with a phase of 0), it is an elementary fact of antenna array theory that the $n$th antenna will experience a phase of $\exp\{j2\pi \frac{d}{\lambda}(n-1)\sin(\theta)\}$ [40], where $\lambda = c/f_c$ is the carrier wavelength.

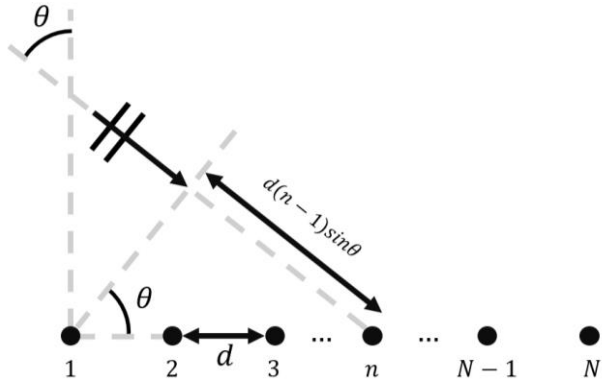

Figure 4.1 A ULA standing in the wavefront of a plane wave, at an angle of $\theta$ with the wavefront. Since the wavefront travels a distance of $d(n-1)\sin\theta$ from the first antenna to the $n$th antenna, there is a phase difference of $e^{j2\pi\frac{d}{\lambda}(n-1)\sin(\theta)}$ between the two antennas.

Since the wave under concern is a plane wave, it can be assumed that there are no other distortions on the signal (such as phase shifts or attenuation) and therefore the channel experienced by each antenna is fully correlated with the channel experienced by the first antenna. In particular, if the first antenna experiences a channel $h_1$, the channel experienced by the $n$th antenna is given by

$$h_n = h_1 \exp\{j2\pi \frac{d}{\lambda}(n-1)\sin(\theta)\}.$$

Then, the covariance of any two channels $h_m$ and $h_n$ can be expressed as

$$E\{h_m h_n^*\} = E\left\{h_1 e^{j2\pi\frac{d}{\lambda}(m-1)\sin(\theta)} h_1^* e^{-j2\pi\frac{d}{\lambda}(n-1)\sin(\theta)}\right\} = \sigma^2 e^{j2\pi\frac{d}{\lambda}(m-n)\sin(\theta)},$$

where $\sigma^2 = E\{h_1 h_1^*\}$ is the variance of $h_1$.

Then,

$$\boldsymbol{R} = E\{\boldsymbol{hh}^H\} = \begin{bmatrix} E\{h_1 h_1^*\} & E\{h_1 h_2^*\} & \cdots & E\{h_1 h_N^*\} \\ E\{h_2 h_1^*\} & E\{h_2 h_2^*\} & \cdots & E\{h_2 h_N^*\} \\ \vdots & \vdots & \ddots & \vdots \\ E\{h_N h_1^*\} & E\{h_N h_2^*\} & \cdots & E\{h_N h_N^*\} \end{bmatrix}$$

$$= \sigma^2 \begin{bmatrix} 1 & e^{-j2\pi\frac{d}{\lambda}\sin(\theta)} & \cdots & e^{-j2\pi\frac{d}{\lambda}(N-1)\sin(\theta)} \\ e^{j2\pi\frac{d}{\lambda}\sin(\theta)} & 1 & \cdots & e^{-j2\pi\frac{d}{\lambda}(N-2)\sin(\theta)} \\ \vdots & \vdots & \ddots & \vdots \\ e^{j2\pi\frac{d}{\lambda}(N-1)\sin(\theta)} & e^{j2\pi\frac{d}{\lambda}(N-2)\sin(\theta)} & \cdots & 1 \end{bmatrix}$$

$$= \sigma^2 N \boldsymbol{u}(\theta)\boldsymbol{u}^H(\theta),$$

where $\boldsymbol{u}(\theta) = \frac{1}{\sqrt{N}}\left[1, e^{j2\pi\frac{d}{\lambda}\sin(\theta)}, \ldots, e^{j2\pi\frac{d}{\lambda}(N-1)\sin(\theta)}\right]^T$ is the unit-energy steering vector for a ULA with an antenna spacing of $d$.

### 4.3.2    Channel Covariances for Multiple Beams

Now, consider an antenna that stands in the wavefront of $N_b$ plane waves, each carrying the same information, as depicted in Figure 4.2.

Let us assume that each plane wave arrives at the antenna array at the same time instant. Naturally, each plane wave will have experienced a different channel before it arrives at the antenna array and each plane wave will have different amplitudes.

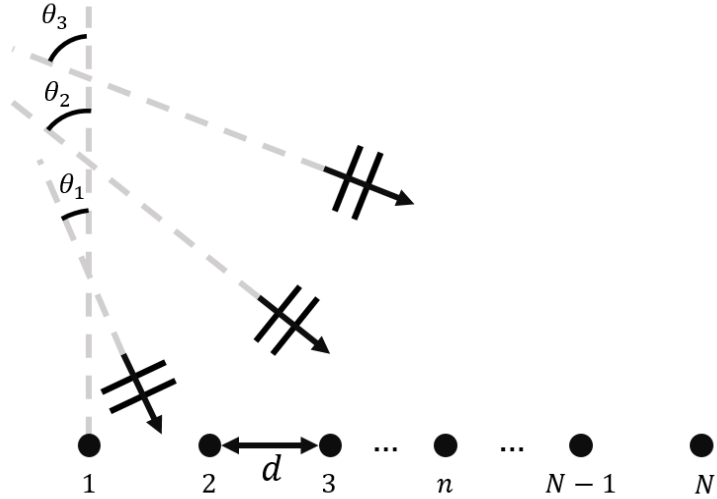

Figure 4.2 A ULA at the wavefront of $N_b = 3$ planar waves, each of which are at a different angle with the antenna array.

Then, the effective channel experienced by the information carried on these waves can be thought of as a combination of the channels experienced by these waves, $h_{1i}$. Let us represent this channel as

$$h_1 = \sum_{i=1}^{N_b} \alpha_i h_{1i},$$

where $\alpha_i$ represents the relative power of each independently distributed, unit-power channel $h_{1i}$.

Now, consider the wave received by the $n$th antenna. Given that the bandwidth of the information being carried on the electromagnetic waves is much smaller than the carrier frequency (which is true for almost all wireless communication systems), it can be assumed that the waves arrive at the $n$th antenna without any difference in the envelope, i.e., in the carried information. The $i$th wavefront arrives at the $n$the antenna with a phase difference of $\exp\{j2\pi\frac{d}{\lambda}(n-1)\sin(\theta_i)\}$. Then, the effective channel experienced by the $n$th antenna can be expressed as

$$h_n = \sum_{i=1}^{N_b} \alpha_i h_{1i} e^{j2\pi\frac{d}{\lambda}(n-1)\sin(\theta_i)}.$$

From this expression, it is easy to deduce that the covariance of $h_1$ and $h_n$ will be equal to

$$\begin{aligned}
Cov\{h_m, h_n\} &= E\{h_m h_n^*\} \\
&= E\left\{\sum_{i=1}^{N_b}\alpha_i h_{1i} \sum_{i=1}^{N_b} \alpha_i^* h_{1i}^* e^{j2\pi\frac{d}{\lambda}(m-n)\sin(\theta_i)}\right\} \\
&= E\left\{\sum_{i=1}^{N_b}\alpha_i \alpha_i^* h_{1i} h_{1i}^* e^{j2\pi\frac{d}{\lambda}(m-n)\sin(\theta_i)}\right\} \\
&+ \underbrace{E\left\{\sum_{\substack{i=1 \\ i\neq j}}^{N_b}\sum_{j=1}^{N_b}\alpha_i \alpha_j^* h_{1i} h_{1j}^* e^{j2\pi\frac{d}{\lambda}(m-n)\sin(\theta_j)}\right\}}_{=0} \\
&= \sum_{i=1}^{N_b} \alpha_i \alpha_i^* E\{h_{1i} h_{1i}^*\} e^{j2\pi\frac{d}{\lambda}(m-n)\sin(\theta_i)} \\
&= \sum_{i=1}^{N_b} |\alpha_i|^2 e^{j2\pi\frac{d}{\lambda}(m-n)\sin(\theta_i)},
\end{aligned}$$

where the second term of the third equality is equal to zero because $h_{1i}$ and $h_{1j}$ are independently distributed, zero-mean random variables.

With some mathematical manipulation, similar to the one done in the previous section, it is not hard to see that the CCM for this antenna array will be

$$\boldsymbol{R} = E\{\boldsymbol{hh}^H\} = \sum_{i=1}^{N_b} |\alpha_i|^2 N \boldsymbol{u}(\theta_i)\boldsymbol{u}^H(\theta_i).$$

### 4.3.3 Channel Covariances for Infinitely Many Beams

Finally, consider infinitely many beams distributed between $\theta_{max}$ and $\theta_{min}$, pictured in Figure 4.3. Let the power of the channel experienced by the infinitesimal beam at angle $\theta_i$ be given by $\rho(\theta_i)\Delta\theta$, where $\rho(\theta)$ is the continuous angular power spectrum of the beams and $\Delta\theta$ represents the width of the infinitesimal angle range. (Note that $\rho(\theta_i)\Delta\theta$ replaces $|\alpha_i|^2$ in the case with finitely many beams.)

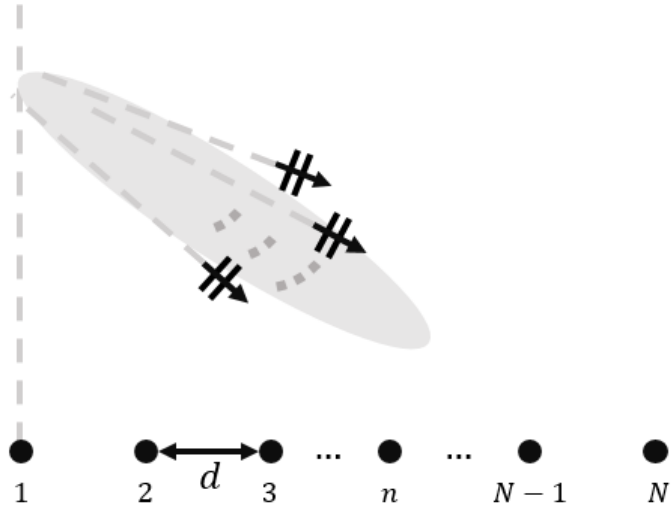

Figure 4.3 A ULA at the wavefront of infinitely many planar waves over a range of angles with the antenna array.

Then, the CCM of the channels experienced by the antennas in the array is given by

$$\boldsymbol{R} = \lim_{N_b \to \infty} \sum_{i=1}^{N_b} (\rho(\theta_i)\Delta\theta) N \boldsymbol{u}(\theta_i) \boldsymbol{u}^H(\theta_i)$$

$$= \int_{\theta_{min}}^{\theta_{max}} \rho(\theta) N \boldsymbol{u}(\theta) \boldsymbol{u}^H(\theta) d\theta.$$

In our case, where the normalized angular power spectrum is assumed to be uniform over the range $[\theta - \frac{\sigma}{2}, \theta + \frac{\sigma}{2}]$, i.e.,

$$\rho(\phi) = \begin{cases} \dfrac{1}{N}\rho, & \phi \in \left[\theta - \dfrac{\sigma}{2}, \theta + \dfrac{\sigma}{2}\right], \\ 0, & otherwise, \end{cases}$$

the CCM is given by

$$\boldsymbol{R} = \rho \int_{\theta-\frac{\sigma}{2}}^{\theta+\frac{\sigma}{2}} \boldsymbol{u}(\phi)\boldsymbol{u}^H(\phi)d\phi.$$

The dummy variable was replaced in the last two equations for the sake of consistency with Eqn. 4.3.

# CHAPTER 5

# DEEP LEARNING AIDED PARAMETRIC CHANNEL COVARIANCE MATRIX ESTIMATION AND HYBRID MASSIVE MIMO BEAMFORMING IN A MILLIMETER-WAVE CHANNEL

In this chapter, the main contributions of this thesis are outlined. First, the system model is described in detail along with the mathematical model used for the channel covariance matrix, and then the estimation method is explained step-by-step.

## 5.1  System Model

### 5.1.1  System Geometry

Consider a system where $K$ single-antenna users are served by one base station (BS) equipped with an $N$-antenna uniform linear array (ULA). For the purpose of CCM estimation, each user transmits a pilot sequence that constitutes of $T$ unit-energy QPSK symbols.

The system operates in a sparse mmWave channel and $L_k$ beams reach the base station for each user $k$. The beams (also called user taps from now on) are uniformly distributed over an angle-delay plane, which spans an angle range between $\theta_{min}$ and $\theta_{max}$ degrees, and a delay range between 0 and $L_{ch} - 1$, where $L_{ch}$ is the number of taps corresponding to the channel delay spread.

Each tap of each user is located at a random place in the angle-delay plane. The angular spread for each user tap is also a random variable, which represents the variance of the AoA, is also a random varible uniformly distributed between $\sigma_{min}$ and $\sigma_{max}$ degrees. A sample visual representation of the AoAs, delays and angular spreads of each user tap in the angle-delay plane is given in Figure 5.1.

The number of active taps for each user is denoted by $L_k$ and is a random integer between 1 and $L_{max}$. The AoA of the $l$th active tap of the $k$th user is denoted by $\theta_k^l$ and is a uniformly distributed random number in the range $[\theta_{min}, \theta_{max}]$. The angular spread of the $l$th active tap of the $k$th user is denoted by $\sigma_k^l$ and is a uniformly distributed random variable in the range $[\sigma_{min}, \sigma_{max}]$. The received SNR of the $l$th active tap of the $k$th user is denoted by $\rho_k^l$ and is a random variable with lognormal distribution. The delay of the $l$th active tap of the $k$th user is denoted by $\tau_k^l$.

With this notation, the system parameters can be expressed in a more formal mathematical language as follows:

$$L_k = \{1, \ldots, L_{max}\} \text{ with prob. } \frac{1}{L_{max}}, \qquad \forall k = 1, \ldots, K,$$

$$\tau_k^l = \{1, \ldots, L_{ch}\} \text{ with prob. } \sim 1/L_{ch}, \qquad \forall l = 1, \ldots, L_k, \forall k = 1, \ldots, K,$$

$$\theta_k^l \sim unif[\theta_{min}, \theta_{max}], \qquad \forall l = 1, \ldots, L_k, \forall k = 1, \ldots, K,$$

$$\sigma_k^l \sim unif[\sigma_{min}, \sigma_{max}] \qquad \forall l = 1, \ldots, L_k, \forall k = 1, \ldots, K,$$

$$10\log_{10}\rho_k^l \sim \mathcal{CN}(\rho_{mean}, \rho_{var}), \qquad \forall l = 1, \ldots, L_k, \forall k = 1, \ldots, K.$$

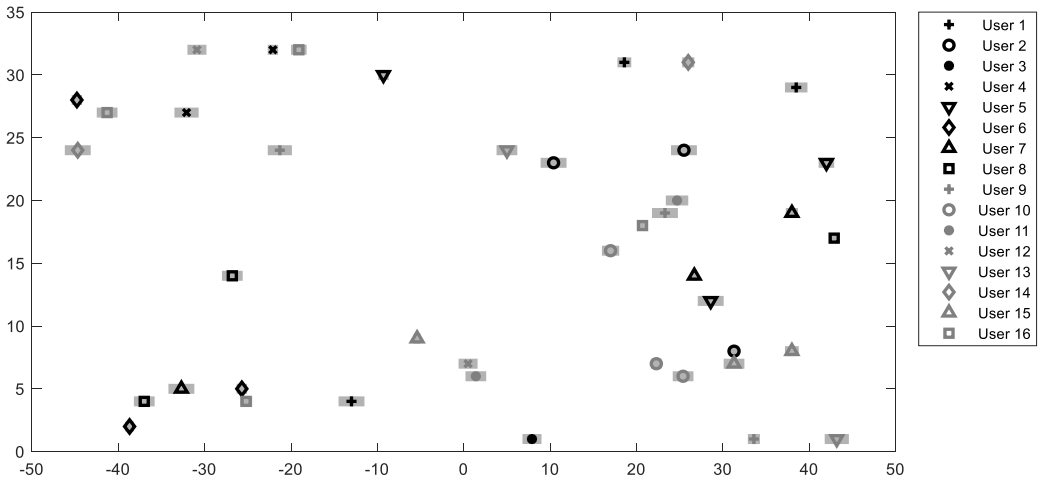

Figure 5.1 A sample angle-delay plane with $K = 16$, $L_{max} = 4$, $L_{ch} = 32$, $\theta_{min} = -45°$, $\theta_{max} = 45°$, $\sigma_{min} = 0.6°$, $\sigma_{max} = 3°$.

Figure 5.1 shows a sample realization for the parameter distributions given above. The AoA for each user tap is represented by a marker, with unique markers for each user. The angular spreads of the user taps are represented with gray rectangles, where each rectangle spans the angle range $[\theta_k^l - \sigma_k^l/2, \theta_k^l + \sigma_k^l/2]$.

In this system model, only the uplink connection is considered. The CCM estimation, instantaneous channel estimation, and data transmission are all done in the uplink. Under TDD operation, all of these can be simply applied to the downlink.

Each tap $l$ of each user $k$ experiences an $N \times 1$ SIMO channel $\boldsymbol{h}_k^l \sim \mathcal{CN}(\boldsymbol{0}, \boldsymbol{R}_k^l)$ that is independent of other taps (both the other taps of the same user and the taps of the other users). In other words, $\boldsymbol{h}_k^l$ is a vector of complex Gaussian distributed random variables with a covariance matrix $\boldsymbol{R}_k^l$, where $\boldsymbol{R}_k^l$ is the channel covariance matrix (CCM) described in Chapter 4.

### 5.1.2 CCM Definition

As stated in the previous section, $\boldsymbol{h}_k^l \sim \mathcal{CN}(\boldsymbol{0}, \boldsymbol{R}_k^l)$, and as stated in the previous chapter, $\boldsymbol{R}_k^l = E\left\{\boldsymbol{h}_k^l (\boldsymbol{h}_k^l)^T\right\}$. $\boldsymbol{R}_k^l$ can be expressed in terms of the angles of arrival (AoA) and the angular spreads of the users as

$$\boldsymbol{R}_k^l = \rho_k^l \int_{\theta_k^l - \frac{\sigma_k^l}{2}}^{\theta_k^l + \frac{\sigma_k^l}{2}} \boldsymbol{u}(\phi) \boldsymbol{u}^H(\phi) d\phi, \tag{5.1}$$

where

$$\boldsymbol{u}(\phi) = \frac{1}{\sqrt{N}} \left[1, e^{j\pi \sin(\phi)}, \dots, e^{j\pi(N-1)\sin(\phi)}\right]^T$$

is the unit-energy steering vector for a ULA. A justification of this mathematical model can be found in Chapter 4.

## 5.2 CCM Estimation

Since $\boldsymbol{R}_k^l$ is uniquely determined by $\theta_k^l$, $\sigma_k^l$ and $\rho_k^l$, it is sufficient to estimate these parameters to accurately determine $\boldsymbol{R}_k^l$ for each active tap $l$ of each user $k$. This is the approach adopted in this thesis, where each of theses three parameters are independently estimated and the CCM estimate $\widehat{\boldsymbol{R}}_k^l$ is determined from the estimated parameters as

$$\widehat{\boldsymbol{R}}_k^l = \hat{\rho}_k^l \int_{\hat{\theta}_k^l - \frac{\hat{\sigma}_k^l}{2}}^{\hat{\theta}_k^l + \frac{\hat{\sigma}_k^l}{2}} \boldsymbol{u}(\phi) \boldsymbol{u}^H(\phi) d\phi.$$

The channel $\boldsymbol{h}_k^l$ is assumed to be a fast-time parameter that changes every frame, whereas the CCM, $\boldsymbol{R}_k^l$, is a slow-time parameter that only needs to be estimated every few frames.

For the purpose of CCM estimation, each user transmits a unique pilot of $T$ unit-energy QPSK symbols in $T_r$ different coherence bandwidths. These pilots have been randomly selected and fixed and are not necessarily orthogonal. No attention was paid to pilot selection or orthogonality since the pilots can arrive at different delays, which makes any assumption of orthogonality invalid.

Under these conditions, the received signal at the BS, $\boldsymbol{Y}$, that corresponds to one pilot sequence can be expressed as

$$\boldsymbol{Y}_{N \times T} = \sum_{k=1}^{K} \boldsymbol{h}_k^l \boldsymbol{x}_k + \boldsymbol{n}_{N \times T},$$

where $\boldsymbol{x}_k = [x_{k,1} \; x_{k,2} \; ... \; x_{k,T}]$ is the $1 \times T$ pilot sequence for the $k$-th user[3], $\boldsymbol{h}_k^l$ is the insantaneous SIMO channel of the $l$th tap of the $k$th user, and the matrix $\boldsymbol{n}$ represents spatially white additive white gaussian noise (AWGN), with $n_{i,j} \sim \mathcal{CN}(0, N_0)$ $\forall i \in \{1, \ldots, N\}$, $j \in \{1, \ldots, T\}$.

---

[3] The reason why $\boldsymbol{x}_k$ are denoted by row vectors rather than the conventional column vectors is the existence of two dimensions in the received signal $\boldsymbol{Y}$. The vertical dimension represents the spatial axis, i.e. the different antennae, and the horizontal dimension represents the temporal axis.

The estimation of each of the parameters $\theta_k^l$, $\sigma_k^l$ and $\rho_k^l$ is described separately in the following sections.

### 5.2.1 AoA Estimation

The estimation of the AoAs is done with the help of a DNN, but the received signal undergoes some preprocessing before it is fed to the DNN. The steps of this preprocessing are:

- User detection, where the coarse position of each user is determined, (This information is assumed to be perfectly available to the CCM estimator.)
- Projection onto the DFT domain, where the dimensios of the received signal are reduced to enable simplified computation,
- Matched filtering/adaptive matched filtering, where the channels of each user are isolated through correlation with the pilot sequences,
- AoA estimation with DNN, where the angle estimate is obtained from the preprocessed data with the help of deep neural networks.

Each step is described in detail in the sections below.

#### 5.2.1.1 User Detection

In this thesis, the angle-delay plane was assumed to be located in the angle range between $-45°$ and $45°$ and the delay spread of the channel was assumed to be 32 taps.

Because there are a limited number of RF chains and limited computational capabilities available to the BS, it is not feasible to perform the CCM estimation simultaneously for every user in the angle-delay plane. For this reason, the angle range in the system geometry was divided into $N_{sec}$ sectors of $90°/N_{sec}$ degree width and each delay tap was processed separately.

In other words, it was assumed that user detection was performed before the CCM estimation stage. The user detection stage detects the user taps in each angle-delay sector and the CCM estimator separately performs the estimations in each sector. The angle-delay sectors defined for the case of $N_{sec} = 8$ can be seen in Figure 5.2. The angle-delay sector for the angle range $[0°, 11.25°]$ was outlined with bold black lines.

Since the main problem considered in this thesis is CCM estimation, it is assumed that the CCM estimator has perfect knowledge of the sectors for each user tap.

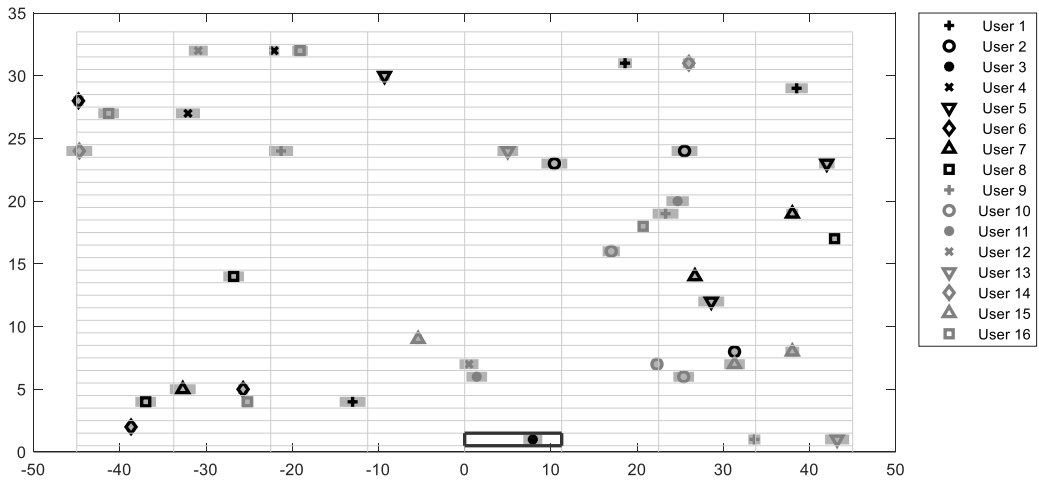

Figure 5.2 The angle-delay sectors found in the angle-delay plane.

### 5.2.1.2  Projection onto the DFT Domain

After matched filtering, an $N \times 1$ received signal vector is obtained for each user. For a massive MIMO scenario, $N$ is a large number, which significantly complicates the CCM estimation. Also, the CCM is not a full-rank matrix, which means that a smaller number of dimensions is enough to represent and estimate it.

For these reasons, the received signal vector is projected into the angle domain before the AoA of the corresponding user is determined. This is done with the help beamforming vectors. The projection of a vector $\boldsymbol{y}_k$ on an angle $\phi_i$ can be obtained by taking its dot product with the beamforming vector corresponding to that angle,

$$b_{i,k} = \boldsymbol{u}^H(\phi_i)\boldsymbol{y}_k,$$

where $\boldsymbol{u}(\phi_i)$ is as defined in Eqn. 2.

The angular power profile $\boldsymbol{b}_k$ of the $k$-th user over a range of angles $\Phi = \{\phi_1, \phi_2, \ldots, \phi_P\}$ can be obtained by finding the projection of the vector $\boldsymbol{y}_k$ over each of these angles.

$$\boldsymbol{b}_k = \begin{bmatrix} b_{1,k} \\ b_{2,k} \\ \vdots \\ b_{P,k} \end{bmatrix} = \begin{bmatrix} \boldsymbol{u}^H(\phi_1) \\ \boldsymbol{u}^H(\phi_2) \\ \vdots \\ \boldsymbol{u}^H(\phi_P) \end{bmatrix} \boldsymbol{y}_k = \boldsymbol{U}^H(\Phi)\boldsymbol{y}_k \quad (5.2)$$

One beam set $\Phi$ was selected for each $90°/N_{sec}$ angle range. For each angle range, eight beams were selected, i.e., $P = N_{sec}$ for each angle sector. For simplicity, the beamforming vectors were selected among the columns of the $N \times N$ DFT matrix, which have the form of a beamforming vector for specific angles.

#### 5.2.1.2.1 The DFT Matrix

The $N \times N$ DFT matrix $\boldsymbol{W}$ is defined elementwise as

$$W_{m,n} = \frac{\omega^{(m-1)(n-1)}}{\sqrt{N}}, \quad \text{where } \omega \triangleq e^{-j\frac{2\pi}{N}}.$$

In explicit form, the DFT matrix can be expressed as

$$\boldsymbol{W} = \frac{1}{\sqrt{N}} \begin{bmatrix} 1 & 1 & 1 & \cdots & 1 \\ 1 & \omega & \omega^2 & \cdots & \omega^{N-1} \\ 1 & \omega^2 & \omega^4 & \cdots & \omega^{2(N-1)} \\ \vdots & \vdots & \vdots & \ddots & \vdots \\ 1 & \omega^{N-1} & \omega^{(N-1)} & \cdots & \omega^{(N-1)(N-1)} \end{bmatrix}.$$

Consider the $n$th column of $\mathbf{W}$,

$$\mathbf{w}_n = \frac{1}{\sqrt{N}}\begin{bmatrix} 1 \\ \omega^n \\ \omega^{2n} \\ \vdots \\ \omega^{(N-1)n} \end{bmatrix} = \frac{1}{\sqrt{N}}\begin{bmatrix} 1 \\ e^{-j\frac{2\pi}{N}n} \\ e^{-j\frac{2\pi}{N}n\cdot 2} \\ \vdots \\ e^{-j\frac{2\pi}{N}n\cdot(N-1)} \end{bmatrix}.$$

Now, consider the steering vector for a ULA with antenna spacing $d = \lambda/2$,

$$\mathbf{u}(\phi) = \frac{1}{\sqrt{N}}\begin{bmatrix} 1 \\ e^{j\pi\sin(\phi)} \\ e^{j\pi\sin(\phi)\cdot 2} \\ \vdots \\ e^{j\pi\sin(\phi)\cdot(N-1)} \end{bmatrix}.$$

It is not hard to see that the $n$th column of $\mathbf{W}$ is equal to the steering vector $\mathbf{u}(\phi)$ if

$$e^{j\pi\sin(\phi)} = e^{-j\frac{2\pi}{N}n}.$$

For each $n$, there is a unique value of $\phi \in \left[-\frac{\pi}{2}, \frac{\pi}{2}\right)$ whose corresponding steering vector is equal to $\mathbf{w}_n$. However, these angles are not uniformly distributed over this range. This nonuniformity is particularly apparent for angles outside $[-45°, 45°]$, but it is significant for the whole angle range. For example, $|\phi_2 - \phi_1| = 0.895°$ and $|\phi_{40} - \phi_{39}| = 1.12°$. There is a very significant variance among the gaps between DFT beams for different angle ranges.

Due to the nonuniformity of these beams in the angle domain, the beams for each sector were selected manually. The selection was made in a way that ensures that the whole angle range is covered by the beams.

In this thesis, the cases $N_{sec} = 4$ and $N_{sec} = 8$ were considered. Visualizations of the selected beams and the angle sectors for both cases are given in

Figure 5.3 and the column indices and the angles corresponding to the beams selected for each angle range are given in Table 5.1.

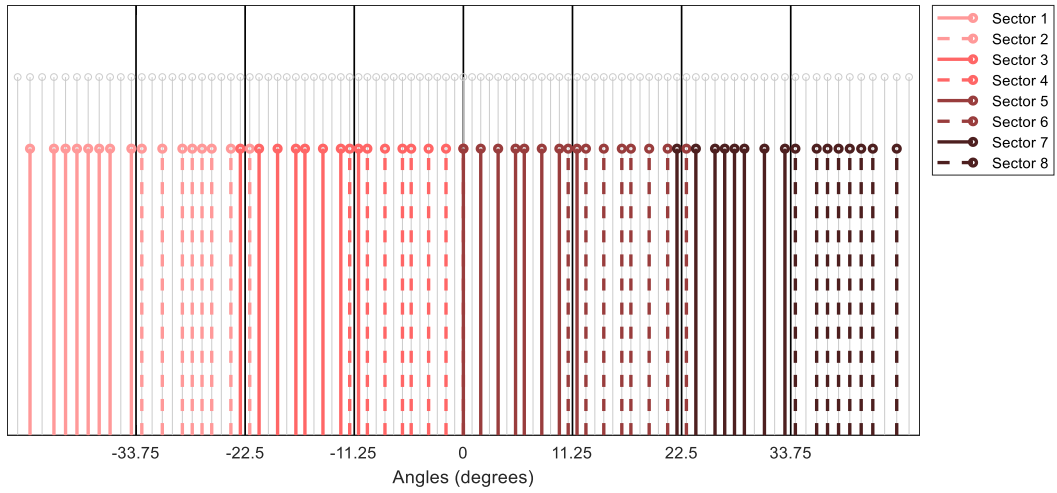

Figure 5.3 The DFT beams that were selected to represent each angle sector ($N_{sec} = 8$).

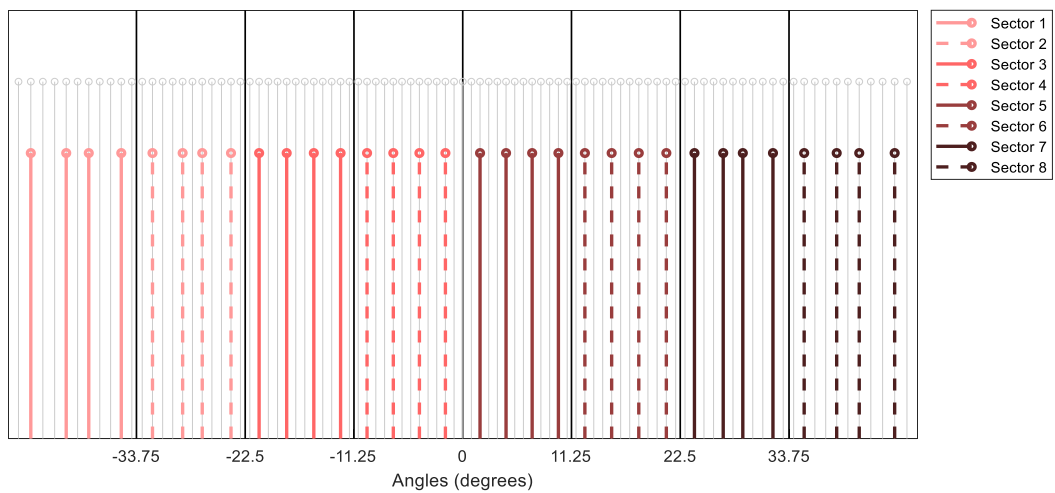

Figure 5.4 The DFT beams that were selected to represent each angle sector ($N_{sec} = 4$).

Table 5.1 Beam indices and angles for each angle sector

| Sector Nr./Range | Beam Column Index/Corresponding Angle in Degrees | | | | | | | |
|---|---|---|---|---|---|---|---|---|
| Sector 1<br>$[-45°, -33.75°]$ | 37 | 39 | 40 | 41 | 42 | 43 | 44 | 46 |
| | -34.2 | -36.4 | -37.5 | -38.7 | -39.8 | -41.0 | -42.2 | -44.7 |
| | 38 | | 41 | | 43 | | 46 | |
| | -35.3 | | -38.7 | | -41.0 | | -44.7 | |
| Sector 2<br>$[-33.75°, -22.5°]$ | 26 | 27 | 29 | 30 | 31 | 32 | 34 | 36 |
| | -22.0 | -24.0 | -25.9 | -26.9 | -28.0 | -29.0 | -31.0 | -33.2 |
| | 27 | | 30 | | 32 | | 35 | |
| | -24.0 | | -26.9 | | -29.0 | | -32.1 | |
| Sector 3<br>$[-22.5°, -11.25°]$ | 14 | 15 | 17 | 19 | 20 | 22 | 24 | 25 |
| | -10.8 | -12.6 | -14.5 | -16.3 | -17.3 | -19.2 | -21.1 | -23.0 |
| | 15 | | 18 | | 21 | | 24 | |
| | -12.6 | | -15.4 | | -18.2 | | -21.1 | |
| Sector 4<br>$[-11.25°, 0°]$ | 1 | 3 | 5 | 7 | 8 | 10 | 12 | 13 |
| | 0.0 | -1.8 | -3.6 | -5.4 | -6.3 | -8.1 | -9.9 | -11.7 |
| | 3 | | 6 | | 9 | | 12 | |
| | -1.8 | | -4.5 | | -7.2 | | -9.9 | |
| Sector 5<br>$[0°, 11.25°]$ | 1 | 127 | 125 | 123 | 122 | 120 | 118 | 117 |
| | 0.0 | 1.8 | 3.6 | 5.4 | 6.3 | 8.1 | 9.9 | 11.7 |
| | 127 | | 124 | | 121 | | 118 | |
| | 1.8 | | 4.5 | | 7.2 | | 9.9 | |
| Sector 6<br>$[11.25°, 22.5°]$ | 116 | 115 | 113 | 111 | 110 | 108 | 106 | 105 |
| | 10.8 | 12.6 | 14.5 | 16.3 | 17.3 | 19.2 | 21.1 | 23.0 |
| | 115 | | 112 | | 109 | | 106 | |
| | 12.6 | | 15.4 | | 18.2 | | 21.1 | |
| Sector 7<br>$[22.5°, 33.75°]$ | 104 | 103 | 101 | 100 | 99 | 98 | 96 | 94 |
| | 22.0 | 24.0 | 25.9 | 26.9 | 28.0 | 29.0 | 31.0 | 33.2 |
| | 103 | | 100 | | 98 | | 95 | |
| | 24.0 | | 26.9 | | 29.0 | | 32.1 | |
| Sector 8<br>$[33.75°, 45°]$ | 93 | 91 | 90 | 89 | 88 | 87 | 86 | 84 |
| | 34.2 | 36.4 | 37.5 | 38.7 | 39.8 | 41.0 | 42.2 | 44.7 |
| | 92 | | 89 | | 87 | | 84 | |
| | 35.3 | | 38.7 | | 41.0 | | 44.7 | |

### 5.2.1.3   Matched Filtering

To isolate the received signals of each user from the signals of the others, the received signal is passed through a bank of matched filters. For an $N \times T$ received signal matrix $Y$, the matched filter output for the $k$-th user is given by

$$\begin{aligned}
y_k &= \frac{1}{\sqrt{T}} Y x_k^H \\
&= \frac{1}{\sqrt{T}} \left( \sum_{i=1}^{K} h_i x_i + n \right) x_k^H \\
&= \frac{1}{\sqrt{T}} \sum_{i=1}^{K} h_i x_i x_k^H + \frac{1}{\sqrt{T}} n x_k^H \\
&= \frac{1}{\sqrt{T}} h_k x_i x_k^H + \frac{1}{\sqrt{T}} \sum_{\substack{i=1 \\ i \neq k}}^{K} h_i x_i x_k^H + \tilde{n} \\
&= \sqrt{T} h_k + \frac{1}{\sqrt{T}} \sum_{\substack{i=1 \\ i \neq k}}^{K} h_i x_i x_k^H + \tilde{n}.
\end{aligned}$$

Since the pilots of the users are not orthogonal, the second term in the last equality is not equal to zero and the received signal of each user has some interference. For larger values of $T$, the level of this interference is not very high, but it still has a significant impact on the CCM estimation performance.

### 5.2.1.4   Angle of Arrival Estimation with Deep Learning

The angle of arrival (AoA) is obtained from the projected signal powers vector $b_k$, which was defined in Eqn. (5.2), with the help of a deep neural network (DNN). It is important to note that $b_k$ is a complex vector, and the squared magnitudes of its elements were fed to the DNN instead of the values themselves, because whereas complex operation is possible for DNNs, it is not very elegant or optimal. The basic principles of DNNs and the mathematical details of the training procedure were explained in Chapter 3. A representation of the AoA estimation step is given in Figure 5.5.

Since there is an asymmetry in the geometry of the beams for different angle sectors, different DNNs are used for each angle sector. No assumption was made about the similarity of the architectures for different angle sectors. In other words, for the DNN used in each angle sector, the network parameters were selected independently of the other sectors.

Of course, the number of layers, the number of nodes and the activation functions that will yield the best performance for this problem are not obvious. Therefore, many neural networks with a variety of combinations of architectures were 'tried out' before the DNN used in this step was selected. The specific configurations that were tried out were numerous and irrelevant, therefore they are not given here.

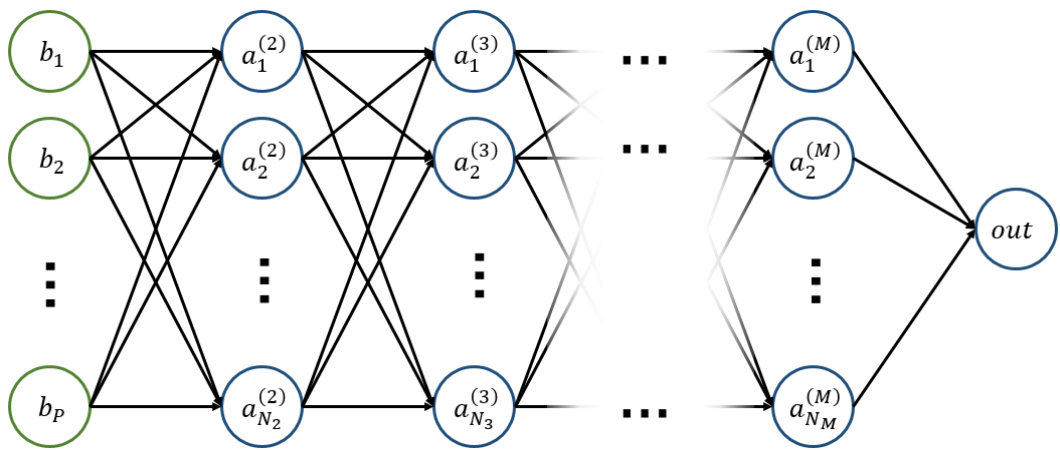

Figure 5.5 AoA estimation with a deep neural network.

The parameters for the selected neural network architectures are given in Appendix A and the performances obtained with these architectures are given in the next chapter.

### 5.2.1.5    Elimination of Outliers

As will be discussed in the next section, information about a single channel realization is not sufficient to estimate the angular spread of a channel. Therefore, all of the procedures described above are repeated $T_r$ times, each time in a different coherence bandwidth. This fact can be exploited to improve the performance of the AoA estimator.

For the estimation of each CCM, $T_r$ different AoA estimations were found, one for each fast-time channel realization. Then, the AoA estimate $\hat{\theta}$ was calculated as the median of these values. The median was selected instead of the mean in order to increase the robustness of the estimate against outliers.

### 5.2.2    Angular Spread Estimation

The second part of parametric CCM estimation is the estimation of the angular spread, which is denoted by $\sigma$ in the CCM expression given in Eqn. (5.1). The angular spread (AS) is not the measure of how wide a beam is for a given channel realization but rather the measure of how much the position of the peak of the angular power spectrum (PAS) varies around its mean value, $\theta$. When the AS of a channel is high, the 'center angle' of independent channel realizations (for a given AoA) varies more around the given AoA. A representation of two channels with relatively low and relatively high angular spreads is given in Figure 5.6. In the figure, both channels have a long term AoA of 0°, but the AS of one of the channels is low, whereas the AS of the other channel is high. It can be seen that for the channel with the higher AS, there is more variation among the angular spectra of independent channel realizations.

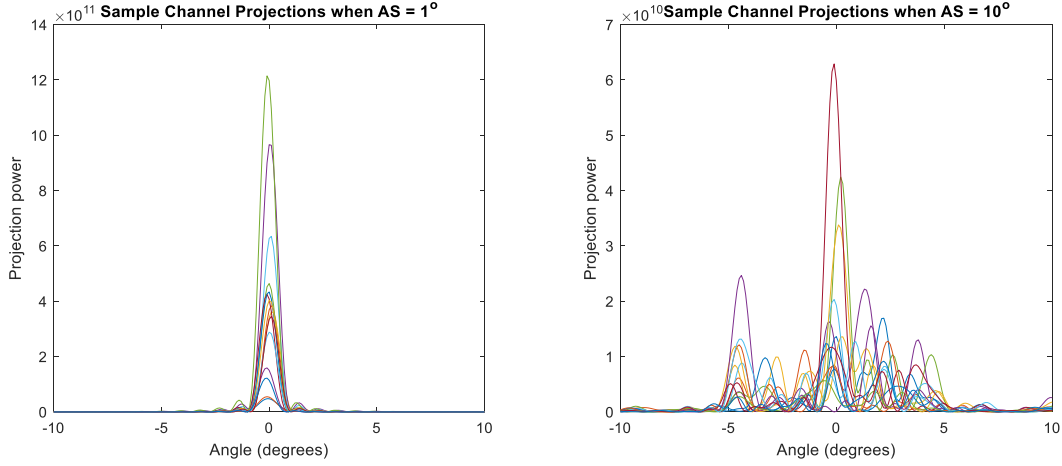

Figure 5.6 Angular power spectra of fifteen independent power realizations with $\sigma = 1$ (left) and $\sigma = 10$ (right).

The estimation of the angular spread is also done with DNNs but a different preprocessing is applied on the data before it is fed to the DNN.

### 5.2.2.1 DFT Beam Selection

Since the angular spread is the measure of how much the peak of the PAS of the channel realization varies around the estimated angle, only the information around the AoA is relevant. Therefore, instead of the full range of the channel projections, only a subset of $N_{sec,2}$ beams, the center of which correspond to the DFT beam with the highest energy, are used for angular spread estimation.

The values of $N_{sec,2}$ that were selected for the two cases studied in this thesis are given in Table 5.1.

Table 5.2 The numbers of DFT beams selected for angular spread estimation, $N_{sec,2}$, for both cases of $N_{sec}$.

| $N_{sec}$ | $N_{sec,2}$ |
|---|---|
| 8 | 5 |
| 4 | 4 |

### 5.2.2.2  Angular Spread Estimation with a DNN

Once the relevant set of channel projections on DFT beams is selected, this dataset is fed to a DNN without any further preprocessing. The DNN operation is practically the same as the AoA estimation case, which is discussed in Section 5.2.1.4 and represented in Figure 5.5. The general description of DNNs can be found in Chapter 3. The performance and the complexity of the DNN-based AS estimator is discussed in the next chapter.

### 5.2.3  Power Estimation

To estimate the third and final parametric component of the CCM, a simple, intuitive method was used. The largest element of the vector $\boldsymbol{b}_k$, which was defined in Eqn. (5.2) was used to estimate the power for the channel,

$$\hat{\rho}_k^l = \max_{i=1,\ldots,P} |b_{k,i}^l|^2,$$

where $b_{k,i}^l$ is the $i$−th element of the vector $\boldsymbol{b}_k^l$. Similar to the AoA estimation, the mean of $T_r$ power estimates was used as the final power estimate in order to increase robustness against statistical outliers.

## 5.3  Beamformer Design with The Estimated CCM

The estimated CCMs can now be used to create relatively isolated physical channels for each user. For this purpose, the estimated CCMs were used for constructing a Capon beamformer. Detailed information about the Capon beamformer, including its derivation and its various implementations, can be found in [41]. The construction of the Capon beamformer for this specific application is described below.

The base station transmits to each user through only one multipath. The multipath with the highest SNR is selected and the transmissions of the other users and also the other multipaths of the same user are considered as interference.

Label all active multipaths of each user with a tap index $t = 1, \ldots, L_{total}$, where

$$L_{total} = \sum_{k=1}^{K} L_k.$$

For each user $k$, denote the index of the multipath with the highest SNR with $t_k$. Then, the Capon beamformer for the $k$−th user, $\boldsymbol{w}^{(k)}$, is given by

$$\boldsymbol{w}^{(k)} = \left(\hat{R}_\eta^{(k)}\right)^{-1} \boldsymbol{h}^{(k)},$$

where $\boldsymbol{h}^{(k)} = \boldsymbol{h}^{(t_k)}$ is the instantaneous channel of the $k$−th user when the $t_k$−th multipath is being used, and

$$\hat{R}_\eta^{(k)} = \sum_{\substack{t'=1 \\ t' \neq t_k}}^{L_k} \hat{R}^{(t')} + N_0 I$$

is the CCM of the interference. In this decription, $\hat{R}^{(k')}$ are the estimated CCMs for each user tap and are used without normalization, i.e., they are scaled according to their respective SNR values. (The instantaneous channels are assumed to be perfectly known in order to avoid any performance loss that might be caused by imperfect knowledge of the channel.)

Then, the detector input $\boldsymbol{y}'^{(k)}$ for the $k$−th user can be obtained from the received signal vector $\boldsymbol{y}^{(k)}$ as

$$\begin{aligned} \boldsymbol{y}'^{(k)} &= \left(\boldsymbol{w}^{(k)}\right)^H \boldsymbol{y}^{(k)} \\ &= \left(\boldsymbol{w}^{(k)}\right)^H Y \boldsymbol{x}_k^H. \end{aligned}$$

When the Capon beamformer is used, the SINR for each user is given by

$$SINR^{(k)} = \frac{\left|\left(\boldsymbol{w}^{(k)}\right)^H \boldsymbol{h}^{(k)}\right|^2 \rho_k}{(\boldsymbol{w}^{(k)})^H R_\eta^{(k)} \boldsymbol{w}^{(k)}},$$

where $\rho_k = \rho_{t_k}$ is the SNR of the strongest multipath of the $k$-th user and $R_\eta^{(k)}$ is the true interference CCM of the $k$-th user (not the estimated interference CCM). The CDFs of the user SINRs for the proposed method and the two benchmark methods are compared in the following chapter.

## 5.4  The Generalized Eigen-Beamformer (GEB) with the Estimated CCM

The GEB is another beamformer, described in detail in [42], that can utilize the statistical information present in the estimated CCM. The GEB is determined as follows:

Solve the generalized eigenvalue problem for the CCM of the relevant tap and the CCM of its interferers,

$$\widehat{\boldsymbol{R}}^{(k)}\boldsymbol{v} = \lambda \widehat{\boldsymbol{R}}_\eta^{(k)}\boldsymbol{v}.$$

Find the largest eigenvalue and the corresponding eigenvector. The corresponding eigenvector is the generalized eigenbeamformer,

$$\boldsymbol{w} = \boldsymbol{v}_1.$$

The GEB has the property of being a single-rank beamformer that utilizes the information regarding the angular spread. Therefore, it is expected to outperform any method that only utilizes the AoA information for beamforming.

# CHAPTER 6

## RESULTS AND DISCUSSION

In this chapter, the performance and the complexity of the method presented in the previous chapter is analyzed. This is done through Monte Carlo simulations and through comparison with existing methods where applicable.

## 6.1    AoA Estimation

The first (and arguably the most important) step of the CCM estimation procedure is the estimation of the angle of arrival (AoA). In this section, the performance and the complexity of the AoA estimator in the CCM estimation procedure is analyzed through Monte Carlo simulations and some calculations. Two existing methods, the MUSIC algorithm and a type of exhaustive search, here called as MaxBeam, are given as benchmarks. The details of these algorithms can be found in Appendix B.

### 6.1.1    Performance

The angle of arrival MSE is defined as

$$MSE \triangleq E\{|\hat{\theta} - \theta|^2\} = E\{|e|^2\}.$$

The MSE performances of the AoA estimator and the two benchmarks with respect to the SNR and the length of the training sequence, $T$, are given in Figure 6.1 and Figure 6.2, respectively.

The MaxBeam algorithm with digital beamforming (DBF), described in Appendix B, represents a lower bound on the mean squared error. It can be seen that the proposed method surpasses the performance of the MUSIC algorithm with hybrid beamforming.

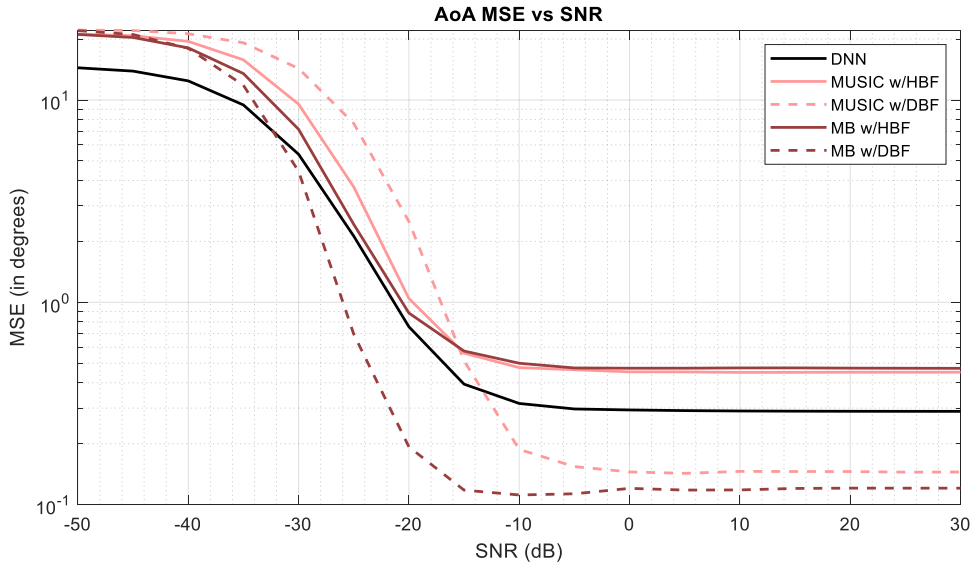

Figure 6.1 AoA MSE vs. mean SNR for different methods, $N_{sec} = 8$. ($T = 128$)

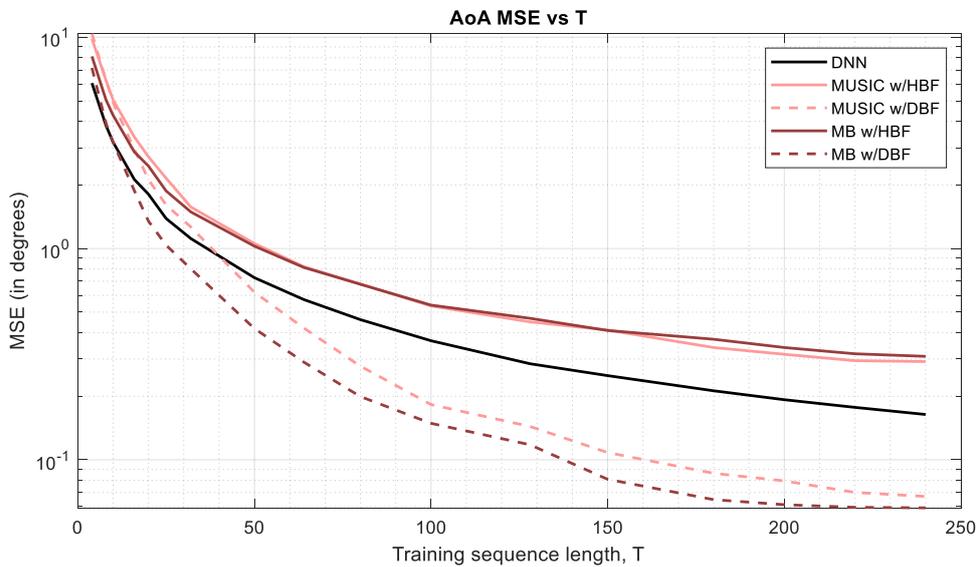

Figure 6.2 AoA MSE vs. training sequence length for different methods, $N_{sec} = 8$. ($Mean\ SNR = 30\ dB$)

It can be seen in the figures that the estimator MSE performance is better than the benchmark methods that use hybrid beamforming. The methods that use digital BF naturally have higher performance, since they have more information to utilize. In practice, digital BF methods are too expensive to be widely used. Therefore, it can be said that the proposed AoA estimation method shows superior MSE performance compared to the benchmark methods with the same amount of available information.

Another measure of performance is the outage probability, defined as the probability that the beam centered around the estimated angle does not cover the true angle of the user. Mathematically expressed,

$$P_{out} \triangleq \Pr\{|\hat{\theta} - \theta| > BW/2\}$$
$$= \Pr\{|e| > BW/2\}.$$

In our case, $BW \sim 360°/N \sim 3°$. Therefore, the $P_{out}$ definition was taken as

$$P_{out} \triangleq \Pr\{|\hat{\theta} - \theta| > 1.5°\}.$$

The outage probabilities of the proposed method and the benchmark methods are given in Figure 6.3 and Figure 6.4. The proposed method has outage probabilities slightly worse than those of the benchmark methods. This is due to the error histogram of the DNN having higher variance compared to the benchmark methods, which can be seen in Figure 6.5. It can be seen that the benchmark methods have 'fatter' histograms that attenuate quickly, whereas the proposed method has a sharp peak that attenuates more slowly, hence the smaller MSE and the higher outage probability.

It can be seen in the following sections that the high outage probability does not degrade the beamforming performance to unacceptable levels.

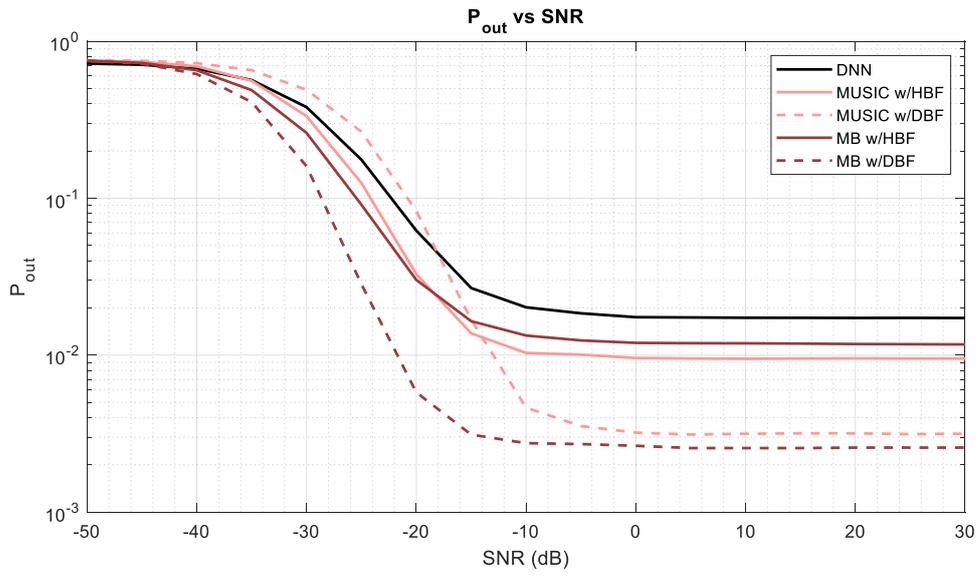

Figure 6.3 $P_{out}$ vs. mean SNR for different methods, $N_{sec} = 8$. ($T = 128$)

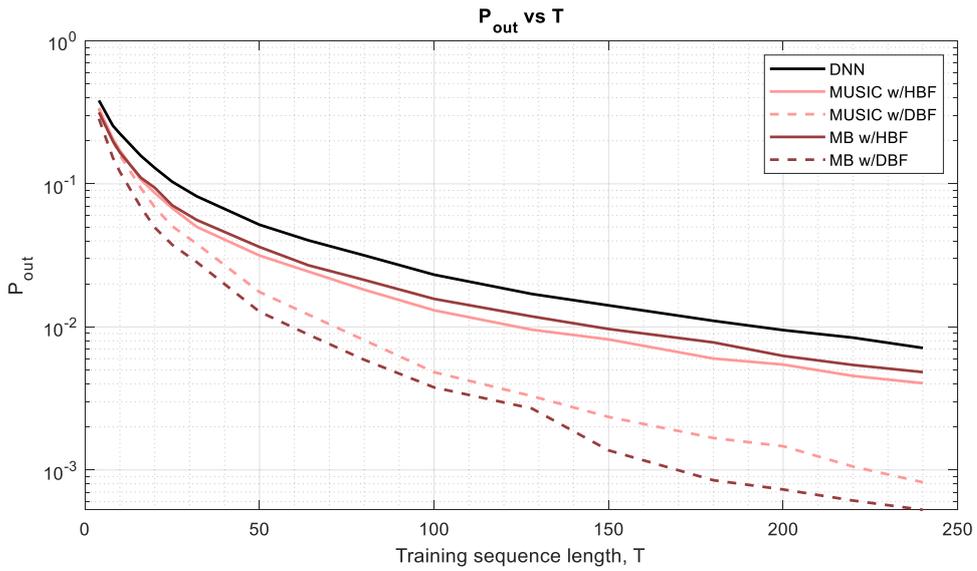

Figure 6.4 $P_{out}$ vs. training sequence length for different methods, $N_{sec} = 8$. ($Mean\ SNR = 30\ dB$)

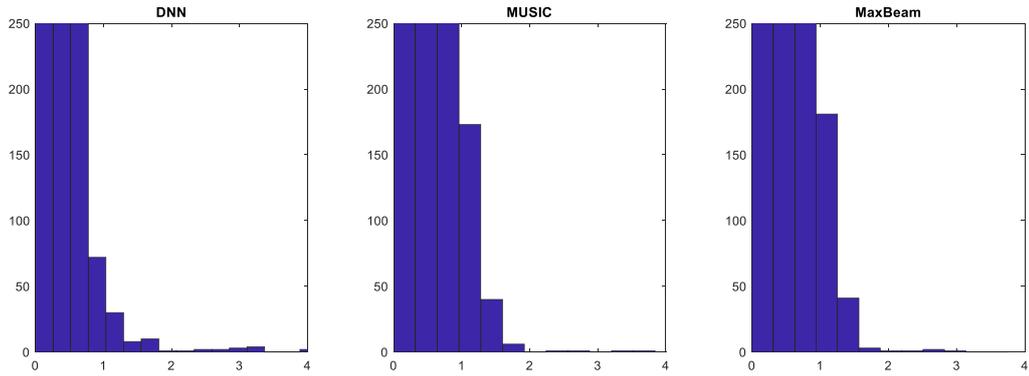

Figure 6.5 Error histograms of the DNN, MUSIC and MaxBeam AoA estimators ($SNR = 30\ dB,\ T = 220, N_{sec} = 8$).

The results given above were also obtained for $N_{sec} = 4$. The obtained results are given in the figures below.

The AoA estimation MSE performances of the proposed method and the benchmarks can be seen in Figure 6.6 and Figure 6.7. It can be seen that the MSE converges to a higher error floor, which is expected since the resolution of the observations is lower. However, as can also be guessed, the complexities and (perhaps more importantly) the number of RF chains required are lower compared to the $N_{sec} = 8$ case.

The outage probabilities of the proposed method and the benchmark methods are given in Figure 6.8 and Figure 6.9. It can be seen that the outage performance of the proposed method is very similar to those of the benchmarks for this case. It can be deduced that the proposed method is more advantageous when the amount of available information is more limited.

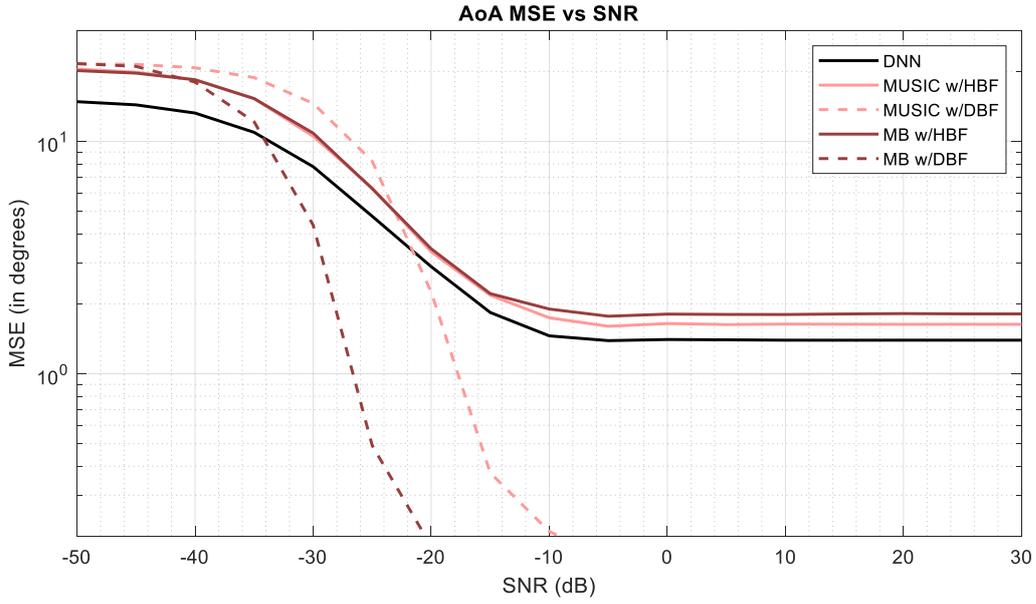

Figure 6.6 AoA MSE vs. mean SNR for different methods, $N_{sec} = 4$. ($T = 128$)

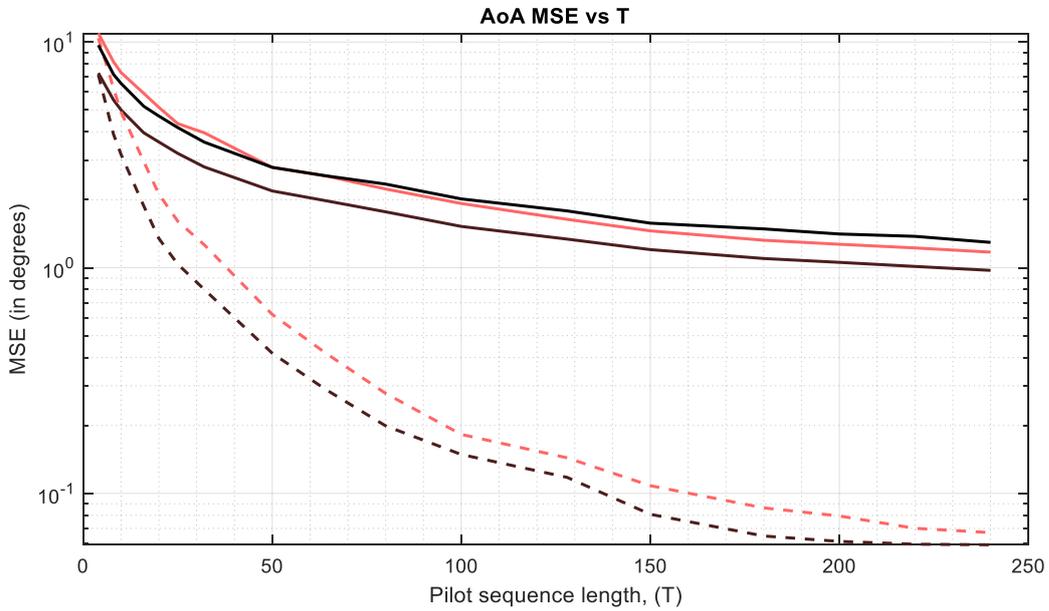

Figure 6.7 AoA MSE vs. training sequence length for different methods, $N_{sec} = 4$. ($Mean\ SNR = 30\ dB$)

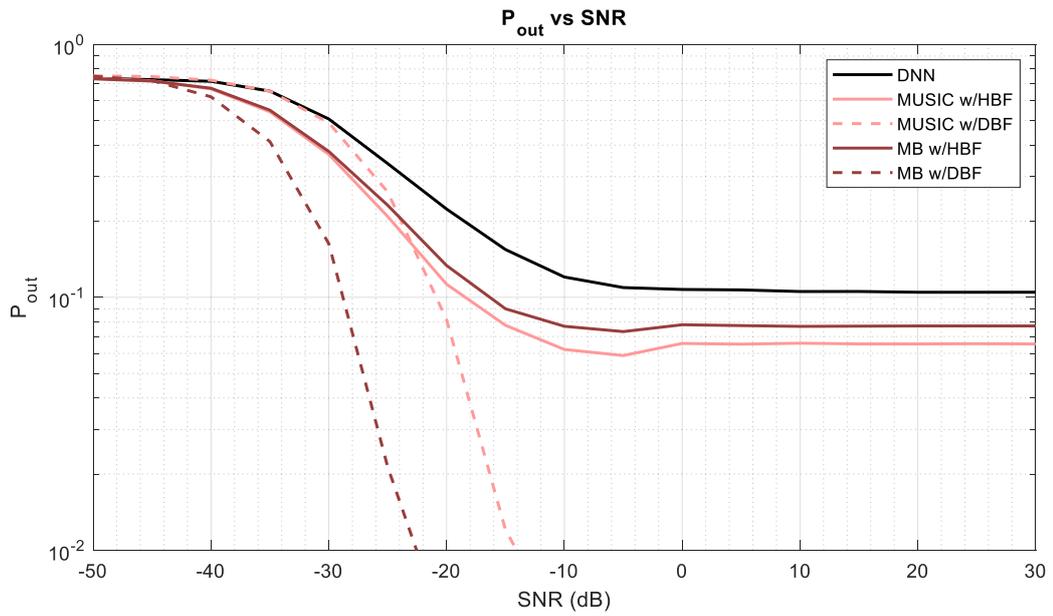

Figure 6.8 $P_{out}$ vs. mean SNR for different methods, $N_{sec} = 4$. ($T = 128$)

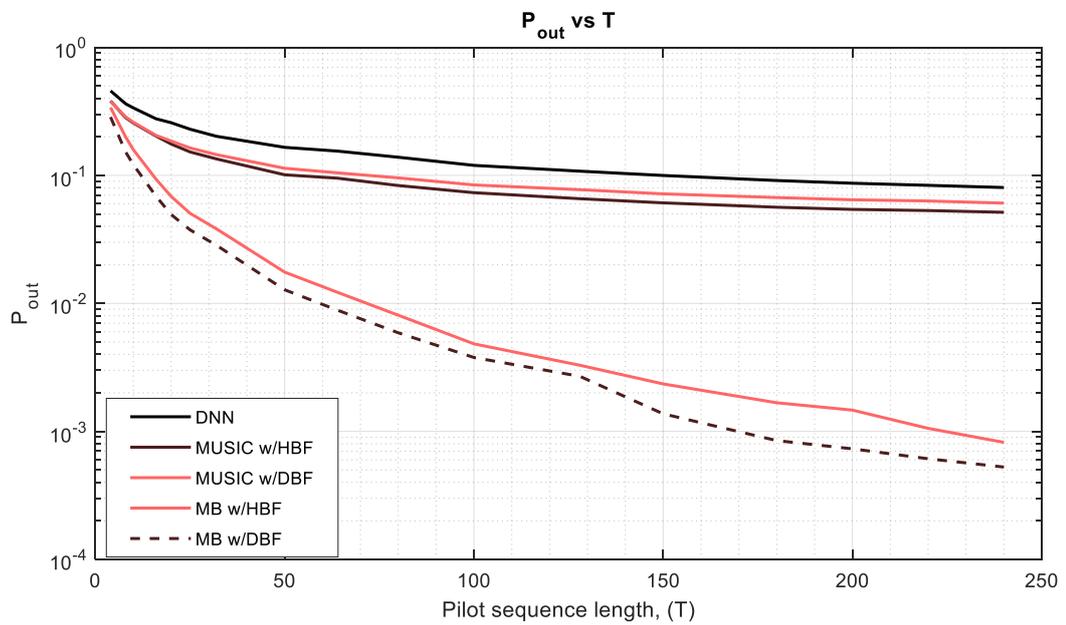

Figure 6.9 $P_{out}$ vs. training sequence length for different methods, $N_{sec} = 4$. ($Mean\ SNR = 30\ dB$)

### 6.1.2 Complexity

The complexity of the proposed AoA estimator was determined in terms of the number of flops, i.e., the number of arithmetic operations needed to obtain the AoA estimate. The Big O notation and the number of flops for each method were used to express the complexities. Details about the Big O notation can be found in [43].

The complexities of the proposed method and the benchmarks are given in Table 6.1. It can be seen in the table that the proposed method has the lowest complexity among all of the given methods. It should be noted that the computational complexity of DNNs rely very heavily on the selected layer sizes, and this comparison may not hold for larger neural networks. However, smaller neural networks have yielded satisfactory performance in this application, making DNNs a desirable candidate for AoA estimation.

Table 6.1 The computational complexities of the methods used in AoA estimation.

| Method | Complexity in Big-O Notation | Complexity in Average Number of Flops | Number of Flops for Sample Case |
|---|---|---|---|
| DNN aided CCM estimator (proposed) | $O(T_r)$ | $T_r N_{DNNnodes}$ | $\sim 2 \cdot 10^3$ |
| MUSIC with HBF | $O(N^3)$ | $NN_{sec}T_r + N^2 T_r + N^3 + N_{grid}N$ | $\sim 2.2 \cdot 10^6$ |
| MUSIC with DBF | $O(N^3)$ | $N^2 T_r + N^3 + N_{grid}N$ | $\sim 2.2 \cdot 10^6$ |
| maxBeam with HBF | $O(NT_r)$ | $NN_{sec}T_r + N_{grid}NT_r$ | $\sim 2 \cdot 10^4$ |
| maxBeam with DBF | $O(NT_r)$ | $N_{grid}NT_r$ | $\sim 1.4 \cdot 10^4$ |

## 6.2 Spread Estimation

The second step of the CCM estimation procedure is the estimation of the angular spread. This information can be used for crafting a more precise beamformer and thereby mitigating inter-user interference.

### 6.2.1 Performance

The main metric used to evaluate the performance of the estimator is the MSE. The MSE performance of the proposed method for $N_{sec} = 4$, $N = 8$ can be seen in Figure 6.10. It can be seen that the AS estimates are pretty robust against low SNR values, and they only start to degrade in performance after the SNR falls below -10dB.

### 6.2.2 Complexity

Similar to the AoA estimator, the complexity of the proposed spread estimator was determined in terms of the number of flops, i.e., the number of arithmetic operations needed to obtain the estimation result. The complexity can be found in Table 6.2.

Table 6.2 The computational complexities of the methods used in AS estimation.

| Method | Complexity in Big-O Notation | Complexity in Average Number of Flops | Number of Flops for Sample Case |
|---|---|---|---|
| Proposed method | $O(T_r)$ | $T_r N_{DNNnodes}$ | $\sim 2 \cdot 10^3$ |

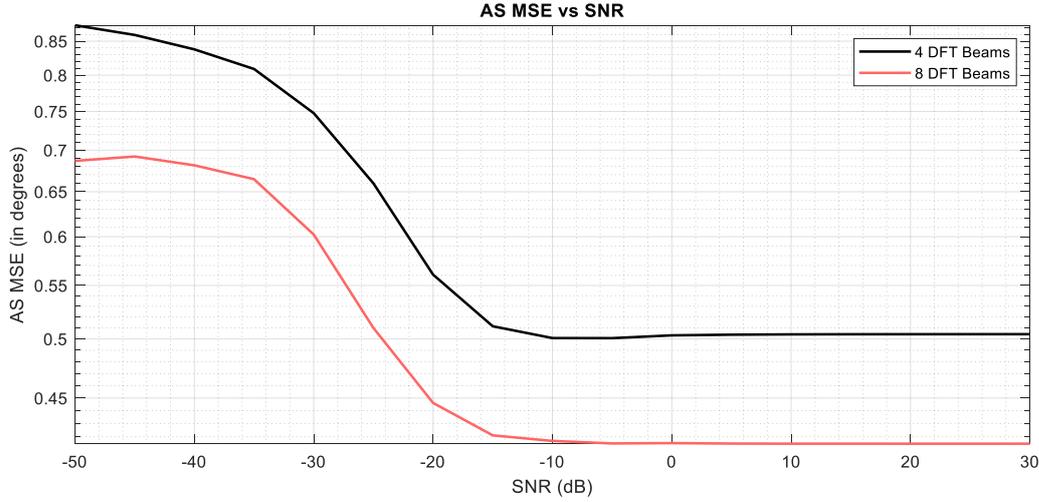

Figure 6.10 AS MSE vs. SNR for $N_{sec} = 4, 8$ ($T = 128$).

## 6.3 Beamforming SINR Performance With Various Methods

In this section, the performance of the proposed method is inspected in terms of SINR CDFs when the estimated CCMs are used to construct Capon beamformers. Since the benchmark methods do not include a method to estimate the angular spread (AS), the method is compared with the case where the true AoA, AS and channel power values are known and are used to construct a perfect CCM. It can be seen in that the proposed method achieves a performance very close to that of the perfect information case.

Another beamforming method, called the Generalized Eigenbeamformer (described in Section 5.4) was also applied. This beamformer is a single-rank beamformer, which has the advantage of having lower complexity but naturally suffers performance-wise. For a fair comparison, the AoA estimates found with the benchmark methods were used to create a steering vector, which served as a single-rank beamformer. The results are significantly worse compared to the Capon beamformer, as can be seen in Figure 6.12, which is expected. The proposed method significantly outperforms the benchmarks, since the GEB uses the AS information whereas the other methods only utilize the AoA estimates for beamforming.

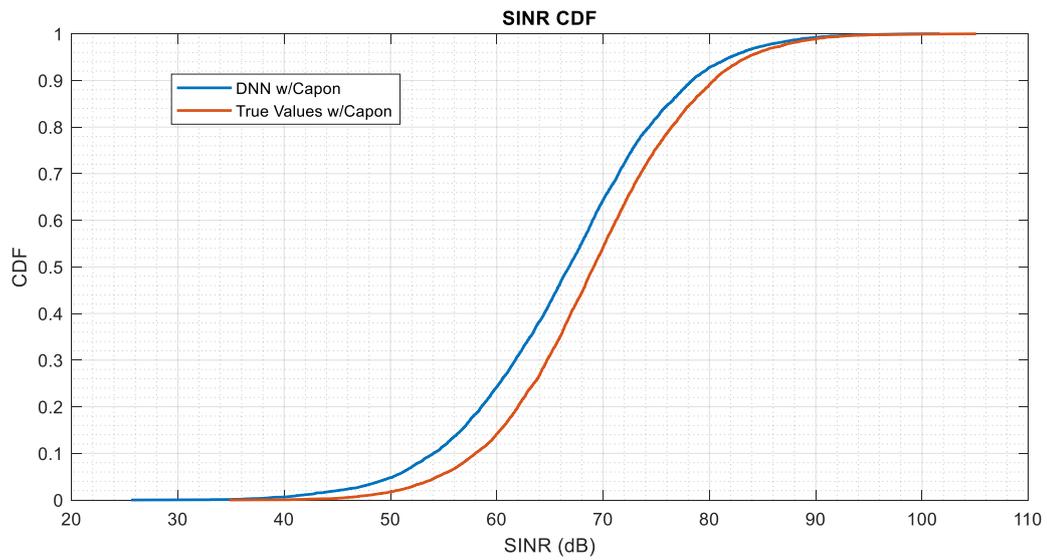

Figure 6.11 The SINR CDFs for the proposed method and the perfect knowledge case. ($N_{sec} = 8, T = 128$)

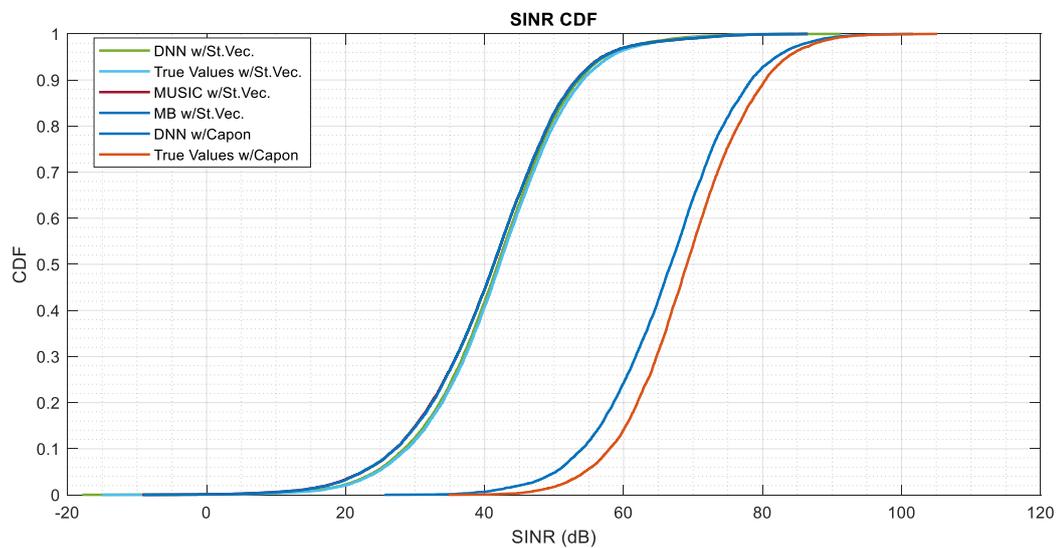

Figure 6.12 Comparison of the Capon beamformer and traditional beam steering with steering vectors. ($N_{sec} = 8, T = 128$)

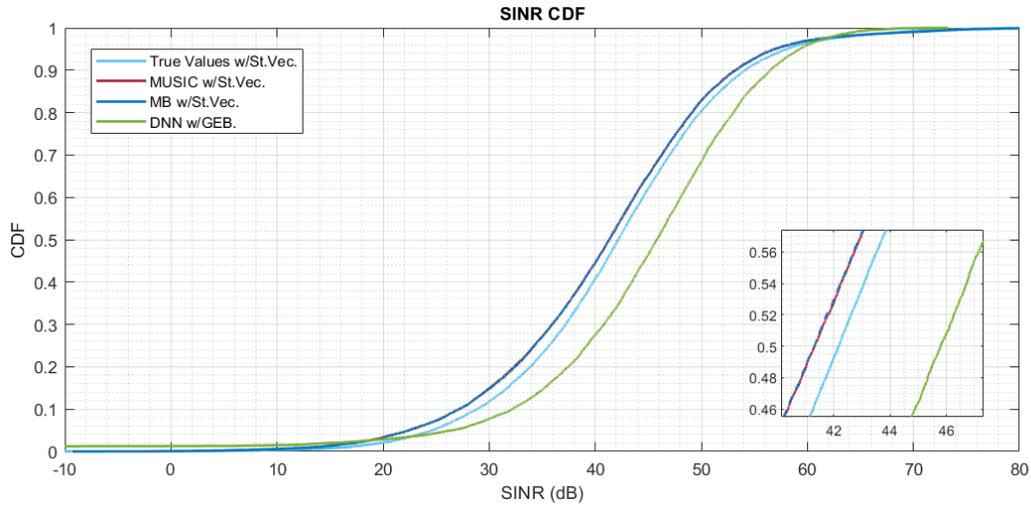

Figure 6.13 Comparison of various methods when used in conjunction with the traditional beamformer. ($N_{sec} = 8, T = 128$)

The results given above were reproduced for the $N_{sec} = 4$ case and can be seen in Figure 6.14, Figure 6.15 and Figure 6.16. It can be seen that the performances behave similarly even when the dimensions of the HBF are reduced.

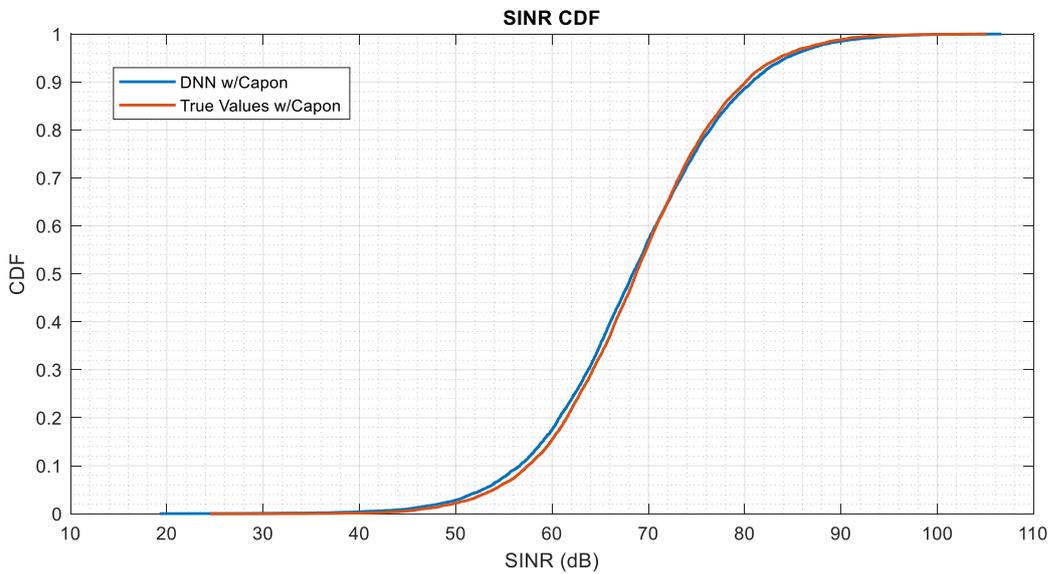

Figure 6.14 The SINR CDFs for the proposed method and the perfect knowledge case. ($N_{sec} = 4, T = 128$)

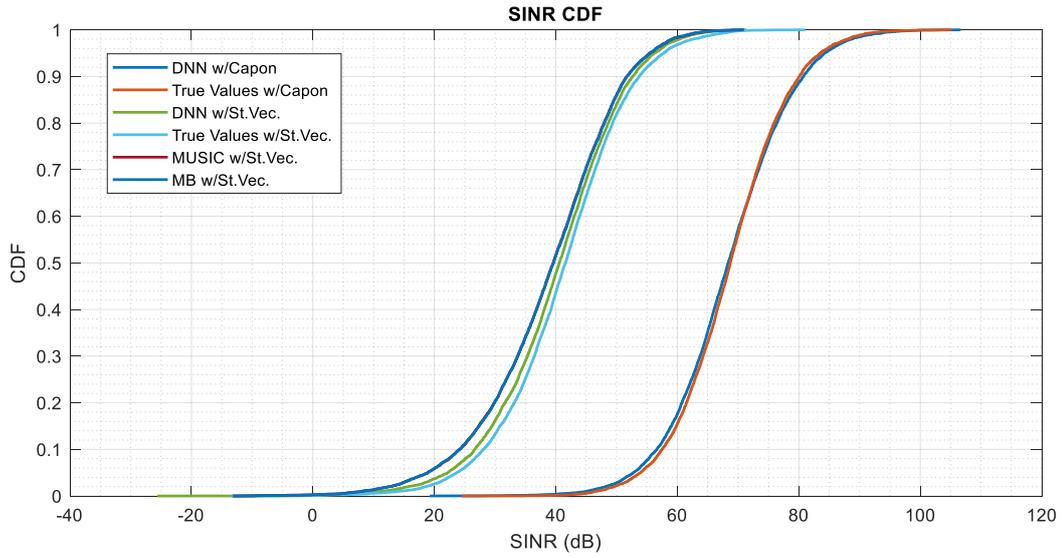

Figure 6.15 Comparison of the Capon beamformer and traditional beam steering with steering vectors. ($N_{sec} = 4, T = 128$)

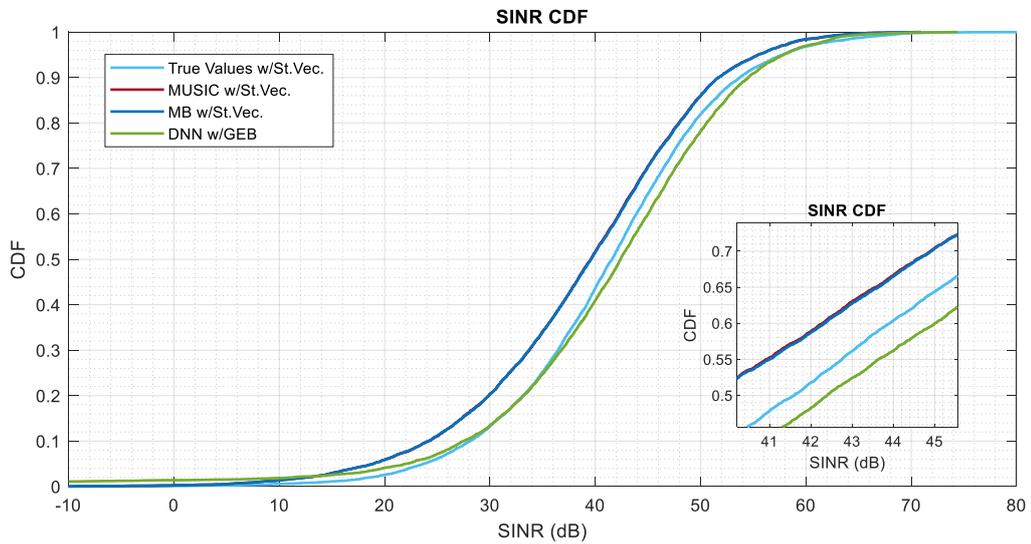

Figure 6.16 Comparison of various methods when used in conjunction with the traditional beamformer. ($N_{sec} = 4, T = 128$)

# CHAPTER 7

# CONCLUSION AND FUTURE WORK

## 7.1 Conclusion

Deep learning methods have attracted a considerable amount of interest in recent years. In this thesis, deep learning methods have been used to solve the problem of channel statistics estimation, which is expected to play an important role in 5G technologies, particularly in the mmWave frequencies. It was observed that the proposed method surpasses the given benchmarks in terms of error performance for the hybrid beamforming case, which is the more realistic scenario. The proposed method was also observed to have comparable or lower computational complexity compared to the benchmarks, which makes the method a desirable candidate for future applications in the estimation of channel statistics. It was observed that the method also yields desirable channel SINR CDF characteristics, when used as the input of a Capon beamformer.

Some further work that can be done to improve this problem is given below:

- Tracking can be utilized to improve the performance of the estimator. Since the positions of the taps will not vary much in adjacent coherence times, tracking is a viable option to improve the estimation performance. In this thesis, it was assumed that channel estimates corresponding to $T_r$ different coherence bandwidths were available to the CCM estimator. In practice, this is not feasible, since it is not acceptable to wait for a few coherence bandwidth before starting data transmission. Instead, $T_r - 1$ previous channel realizations can be kept in memory and the CCM estimate can be updated according to each new channel realization, which would make implementing this estimator much more feasible.

- The user detection stage can be implemented as a classification problem. This stage was assumed to be perfectly carried out. A user detection algorithm can be designed to precede this estimatior. Since the angle sectors and tap delays create a discrete grid over the angle-delay plane, this problem is suitable to be implemented as a classification problem.

# APPENDICES

## A. DNN Architectures

The DNN architectures used in the methods described in this thesis are given below. Some properties are the same for all architectures that have been used in this thesis. These parameters are given in the table below.

Table A.1 Parameters that are common for all the DNN architectures used in this thesis.

| Parameter Name | Parameter Value |
|---|---|
| Activation functions for each hidden layer, $f_2, \ldots, f_M$ | $sig(x) = \dfrac{1}{1 + e^{-x}}$ |
| Activation function for the output layer, $f_{M+1}$ | $f_5(x) = x$ |
| Regularization constant, $r$ | 0 |
| Training function | Gradient descent with momentum |

For the cases $N_{sec} = 4, 8$, the DNN architectrures used for each sector in the AoA estimation step are given in the table below. The same architecture was used for every sector.

Table A.2 Parameters for the architecture used in AoA estimation.

| $N_{sec}$ | Sector No. | Number of Layers, $N_M$ | Number of Nodes in Each Hidden Layer, $\{N_2, \ldots, N_M\}$ |
|---|---|---|---|
| 8 | $1, \ldots, 8$ | 3 | $\{16, 16, 1\}$ |
| 4 | $1, \ldots, 8$ | 3 | $\{16, 16, 1\}$ |

For the case $N_{sec} = 4$, the DNN architectrures used for each sector in the AS estimation step are given in the table below.

Table 7A.3 Parameters for the architecture used in AS estimation. ($N_{sec} = 4$).

| Sector Number | Number of Layers, $N_M$ | Number of Nodes in Each Hidden Layer, $\{N_2, ..., N_M\}$ |
|---|---|---|
| 1 | 4 | $\{32, 32, 32, 1\}$ |
| 2 | 4 | $\{40, 40, 40, 1\}$ |
| 3 | 4 | $\{20, 20, 10, 1\}$ |
| 4 | 4 | $\{40, 20, 10, 1\}$ |
| 5 | 4 | $\{32, 32, 16, 1\}$ |
| 6 | 4 | $\{40, 20, 20, 1\}$ |
| 7 | 4 | $\{40, 20, 10, 1\}$ |
| 8 | 4 | $\{40, 20, 10, 1\}$ |

For the case $N_{sec} = 8$ the DNN architectrures used for each sector in the AS estimation step are given in the table below.

Table A.4 Parameters for the architecture used in AS estimation. ($N_{sec} = 8$).

| Sector Number | Number of Layers, $N_M$ | Number of Nodes in Each Hidden Layer, $\{N_2, ..., N_M\}$ |
|---|---|---|
| 1 | 4 | $\{40, 20, 20, 1\}$ |
| 2 | 4 | $\{40, 20, 20, 1\}$ |
| 3 | 4 | $\{20, 20, 10, 1\}$ |
| 4 | 4 | $\{40, 40, 20, 1\}$ |
| 5 | 4 | $\{32, 16, 16, 1\}$ |
| 6 | 4 | $\{40, 20, 20\ 1\}$ |
| 7 | 4 | $\{40, 40, 20, 1\}$ |
| 8 | 4 | $\{16, 16, 1\}$ |

## B. Detailed Explanation of the MUSIC and the MaxBeam Algorithms

The two main methods that were used for performance comparison with our method were the well-known MUSIC method and a basic method based on angle-domain scanning with steering vectors, which we have named the MaxBeam method.

### B. 1. MUSIC Algorithm with Digital Beamforming

The Multiple Signal Classification (MUSIC) Algorithm is a method based on minimizing the projection of the signal vector on the null space of the channel matrix [44]. It is a widely popular method for frequency estimation and direction finding, specifically when multiple sources are present.

The MUSIC algorithm has the following steps:

1. Obtain an $N \times N$ covariance matrix estimate of the received signal from channel snapshots. (For this scenario, the number of available channel snapshots is the number of channel realizations, $T_r$.)

$$\widehat{R} = \frac{1}{T_r} \sum_{n=1}^{T_r} \widehat{h}_n \widehat{h}_n^T.$$

2. Determine the $N - M$ dimensional null space $V$ of this covariance matrix, where $N$ is the number of antennae and $M$ is the number of signal sources in the environment.

3. Perform an exhaustive search over the range of possible angles $\theta$, find the $M$ values that minimize the projection of their respective steering vectors onto $V$. (In our case, $M = 1$). In other words,

$$\widehat{\theta} = \underset{\theta \in [\theta_{min}, \theta_{max}]}{\operatorname{argmin}} \|V^H u(\theta)\|^2,$$

where $\boldsymbol{u}(\theta) = \frac{1}{\sqrt{N}}\left[1, e^{j\pi \sin(\phi)}, \ldots, e^{j\pi(N-1)\sin(\phi)}\right]^T$ is the $N \times 1$ unit-energy steering vector for a ULA with $\lambda/2$ antenna spacing.

Although the MUSIC algorithm is a very popular method and performs well in specific scenarios, it has some limitations, which are listed below.

- The MUSIC algorithm requires a relatively large number of channel snapshots to yield a good performance. In radar applications this is not a problem, but in our scenario this has presented a disadvantage for the MUSIC algorithm.
- The MUSIC algorithm cannot account for multipaths or variances in the power angular spectrum of the received signals.

## B.2. MUSIC Algorithm with Hybrid Beamforming

In usual radar applications, the MUSIC algorithm is assumed to have measurements from all sensors/antennae. However, in 5G scenarios cost limitations allow for only a small number of RF chains, hence hybrid beamforming must be used. The details of this (projection onto DFT beamspace) are explained in Chapter 5. To summarize, instead of $N$ measurements, only $N_{sec}$ measurements are available to the receiver in the form of

$$\boldsymbol{b}_n = \begin{bmatrix} b_{1,n} \\ b_{2,n} \\ \vdots \\ b_{N_{sec},n} \end{bmatrix} = \begin{bmatrix} \boldsymbol{u}^H(\phi_1) \\ \boldsymbol{u}^H(\phi_2) \\ \vdots \\ \boldsymbol{u}^H(\phi_{N_{sec}}) \end{bmatrix} \boldsymbol{y}_n = \boldsymbol{U}^H(\Phi)\boldsymbol{y}_n,$$

where $\boldsymbol{u}^H(\phi_1)$ correspond to pre-selected columns of the $N \times N$ DFT matrix.

Then, the MF output will not yield an estimate of the channel snapshot, but rather a projection of the channel snapshot onto an $N_{sec}$−dimensional subspace,

$$\boldsymbol{d}_n = \boldsymbol{U}^H(\Phi)\boldsymbol{h}_n.$$

Then, the matrix $\widehat{\boldsymbol{R}}$ must be constructed from the $N_{sec} \times 1$ vectors $\boldsymbol{d}_n$. Is is easy to see that the matrix $\boldsymbol{U}(\Phi)$ is a unitary matrix. Then, it is straigthforward to construct a low-rank approximation $\widetilde{\boldsymbol{h}}_n$ of the $N \times 1$ vector $\boldsymbol{h}_n$ from $\boldsymbol{d}_n$,

$$\widetilde{\boldsymbol{h}}_n = \boldsymbol{U}(\Phi)\boldsymbol{d}_n.$$

With this observation, the autocorrelation matrix can be estimated as

$$\widehat{\boldsymbol{R}} = \frac{1}{T_r} \sum_{n=1}^{T_r} (\boldsymbol{U}(\Phi)\boldsymbol{d}_n)(\boldsymbol{U}(\Phi)\boldsymbol{d}_n)^T.$$

The rest of the algorithm is the same for hybrid and digital BF.

Compared to the MUSIC algorithm with digital BF, the MUSIC algorithm with hybrid BF is a fairer rival to the method proposed in this thesis, since both have the same amount of information available to them.

### B.3.  MaxBeam Method with Digital Beamforming

The method that we have named MaxBeam is a type of exhaustive search (similar to maximum likelihood). Over a grid of angles, the algorithm returns the angle that maximizes the absolute value of the projection of the given channel over the steering vector corresponding to that angle. Mathematically expressed,

$$\hat{\theta} = \underset{\theta \in [\theta_{min}, \theta_{max}]}{\operatorname{argmax}} \widehat{\boldsymbol{h}}^H \boldsymbol{u}(\theta),$$

where $\boldsymbol{u}(\theta) = \frac{1}{\sqrt{N}}\left[1, e^{j\pi \sin(\phi)}, \ldots, e^{j\pi(N-1)\sin(\phi)}\right]^T$ is the $N \times 1$ unit-energy steering vector for a ULA with $\lambda/2$ antenna spacing.

Since $T_r$ snapshots are available to the receiver, the MaxBeam algorithm makes $T_r$ such estimations and takes the median value as the final AoA estimate.

$$\hat{\theta}_n = \underset{\theta \in [\theta_{min}, \theta_{max}]}{\operatorname{argmax}} \widehat{\boldsymbol{h}}_n^H \boldsymbol{u}(\theta),$$

$$\hat{\theta} = median\{\theta_n\}_{n=1}^{Tr}.$$

For a single-user channel with no angular spread, the MaxBeam method would be equivalent to the maximum likelihood method and it would be the optimum solution. However, due to the presence of interferers and angular variance, the MaxBeam method is not an optimum solution. Since the channel under concern is sparse, interference is a rare occurance and this method yields a very good performance.

Even in this relatively simplified form, this algorithm is very complex. Implementing a joint $KL-$dimensional exhaustive search would be prohibitively complex, therefore the MaxBeam algorithm with DBF was used as a lower bound instead of the actual optimum solution.

### B.4. MaxBeam Method with Hybrid Beamforming

Similar to the MUSIC algorithm, an $N \times 1$ input vector for the MaxBeam algorithm can be obtained from the $N_{sec} \times 1$ measurement vector through a straightforward matrix multiplication

$$\tilde{\boldsymbol{h}}_n = \boldsymbol{U}(\Phi)\boldsymbol{d}_n.$$

After $\tilde{\boldsymbol{h}}_n$ has been obtained, the rest of the MaxBeam algorithm is the same as the digital BF case.

Compared to the MaxBeam algorithm with digital BF, the MaxBeam algorithm with hybrid BF is a fairer rival to the method proposed in this thesis, since both have the same amount of information available to them.

# REFERENCES

[1]  E. Björnson, L. V. d. Perre, S. Buzzi and E. G. Larsson, "Massive MIMO in Sub-6 GHz and mmWave: Physical, Practical, and Use-Case Differences," *IEEE Wireless Communications,* vol. 26, no. 2, pp. 100-108, 2019.

[2]  J. Proakis and M. Salehi, Digital Communications, McGraw-Hill, 2001.

[3]  T. Sanguankotchakorn and G. V. Sowmya, "Reducing the Computational Complexity of Massive MIMO using Pre-coding Techniques under Some Lower Orders," *Journal of Communications,* vol. 14, no. 16, pp. 498-503, 2019.

[4]  C. Sun, X. Gao, S. Jin, M. Matthaiou, Z. Ding and C. Xiao, "Beam Division Multiple Access Transmission for Massive MIMO Communications," *IEEE Transactions on Communications,* vol. 63, no. 6, pp. 2170-2184, 2015.

[5]  H. Yin, D. Gesbert, M. Filippou and Y. Liu, "A Coordinated Approach to Channel Estimation in Large-Scale Multiple-Antenna Systems," *IEEE Journal on Selected Areas in Communication,* vol. 31, no. 2, pp. 264-273, 2013.

[6]  H. Xie, F. Gao, S. Jin, J. Fang and Y.-C. Liang, "Channel Estimation for TDD/FDD Massive MIMO Systems With Channel Covariance Computing,"

*IEEE Transactions on Wireless Communications,* vol. 17, no. 6, pp. 4206-4218, 2018.

[7] D. Neumann, M. Joham and W. Utschick, "Covariance Matrix Estimation in Massive MIMO," *arxiv.org,* no. arXiv:1705.02895v2 [cs.IT], 2018.

[8] A. O. Kalayci and G. M. Guvensen, "An Efficient Beam and Channel Acquisition via Sparsity Map and Joint Angle-Delay Power Profile Estimation for Wideband Massive MIMO Systems," *arxiv.org: 1910.05815v1,* 2019.

[9] H. Xie, F. Gao, S. Jin, J. Fang and Y.-C. Liang, "Channel Estimation for TDD/FDD Massive MIMO Systems With Channel Covariance Computing," *IEEE Transactions on Wireless Communications,* vol. 17, no. 6, pp. 4026-4038, 2018.

[10] M. Jordan, A. Dimofte, X. Gong and G. Ascheid, "Conversion from Uplink to Downlink Spatio-Temporal Correlation with Cubic Splines," in *VTC Spring 2009 - IEEE 69th Vehicular Technology Conference*, Barcelona, Spain, 2009.

[11] A. Decurninge, M. Guillaud and D. Slock, "Channel Covariance Estimation in Massive MIMO Frequency Division Duplex Systems," *arxiv.org,* no. arXiv:1602.05741v1 [cs.IT], 2016.

[12] C. Zhang, P. Patras and H. Haddadi, "Deep Learning in Mobile and Wireless Networking: A Survey," *IEEE COMMUNICATIONS SURVEYS & TUTORIALS,* vol. 21, no. 3, pp. 2224-2287, 2019.

[13] K. Zheng and et al., "Big data-driven optimization for mobile networks toward 5G," *IEEE Netw.,* vol. 30, no. 1, pp. 44-51, 2016.*IEEE Transactions on Wireless Communications,* vol. 17, no. 6, pp. 4206-4218, 2018.

[7] D. Neumann, M. Joham and W. Utschick, "Covariance Matrix Estimation in Massive MIMO," *arxiv.org,* no. arXiv:1705.02895v2 [cs.IT], 2018.

[8] A. O. Kalayci and G. M. Guvensen, "An Efficient Beam and Channel Acquisition via Sparsity Map and Joint Angle-Delay Power Profile Estimation for Wideband Massive MIMO Systems," *arxiv.org: 1910.05815v1,* 2019.

[9] H. Xie, F. Gao, S. Jin, J. Fang and Y.-C. Liang, "Channel Estimation for TDD/FDD Massive MIMO Systems With Channel Covariance Computing," *IEEE Transactions on Wireless Communications,* vol. 17, no. 6, pp. 4026-4038, 2018.

[10] M. Jordan, A. Dimofte, X. Gong and G. Ascheid, "Conversion from Uplink to Downlink Spatio-Temporal Correlation with Cubic Splines," in *VTC Spring 2009 - IEEE 69th Vehicular Technology Conference*, Barcelona, Spain, 2009.

[11] A. Decurninge, M. Guillaud and D. Slock, "Channel Covariance Estimation in Massive MIMO Frequency Division Duplex Systems," *arxiv.org,* no. arXiv:1602.05741v1 [cs.IT], 2016.

[12] C. Zhang, P. Patras and H. Haddadi, "Deep Learning in Mobile and Wireless Networking: A Survey," *IEEE COMMUNICATIONS SURVEYS & TUTORIALS,* vol. 21, no. 3, pp. 2224-2287, 2019.

[13] K. Zheng and et al., "Big data-driven optimization for mobile networks toward 5G," *IEEE Netw.,* vol. 30, no. 1, pp. 44-51, 2016.

[14] S. Yu, M. Liu., W. Dou, X. Liu and S. Zhou, "Networking for big data: A Survey," *IEEE Commun. Surveys Tuts.,* vol. 19, no. 1, pp. 531-549, 2017.

[15] Z. C. a. D. B. Smith, "Heterogeneous machine-type communications in cellular networks: Random access optimization by deep reinforcement learning," in *Proc. IEEE Int. Conf. Commun. (ICC),*, 2018.

[16] R. Atallah, C. Assi and M. Khabbaz, "Deep reinforcement learning based scheduling for roadside communication networks," in *Proc. 15th IEEE Int. Symp. Model. Optim. Mobile Ad Hoc Wireless Netw. (WiOpt)*, Paris, France, 2017.

[17] J. Wang et al., "Spatiotemporal modeling and prediction in cellular networks: A big data enabled deep learning approach," in *Proc. 36th Annu. IEEE Int. Conf. Comput. Commun. (INFOCOM)*, Atlanta, GA, USA, 2017.

[18] W. Wang, M. Zhu, J. Wang, X. Zeng and Z. Yang, "End-to-end encrypted traffic classification with one-dimensional convolution neural networks," in *Proc. IEEE Int. Conf. Intell. Security Informat.,*, Beijing, China, 2017.

[19] T. J. O'Shea, T. Erpek and T. C. Clancy, "Deep learning based MIMO communications," in *arXiv preprint arXiv:1707.07980, 2017*, 2017.

[20] N. E. West and T. O'Shea, "Deep architectures for modulation recognition," in *Proc. IEEE Int. Symp. Dyn. Spectr. Access Netw. (DySPAN)*, Piscataway, NJ, USA, 2017.

[21] H. Hongji, Y. Jie, S. Yiwei, H. Hao and G. Guan, "Deep learning for super-resolution channel estimation and DOA estimation based massive MIMO system," *IEEE Trans. Veh. Technol.,* vol. 67, no. 9, pp. 8549-8560, 2018.

[22] L. Bai, C.-X. Wang, J. Huang, Q. Xu, Y. Yang, G. Goussetis, J. Sun and W. Zhang, "Predicting Wireless MmWave Massive MIMO Channel

Characteristics Using Machine Learning Algorithms," *Hindawi Open Access,* vol. 2018, 2018.

[23] L. Cassaccia, "Demystifying 3GPP – An insider's perspective to how 4G and 5G standards get created," 2017. [Online]. Available: https://www.qualcomm.com/news/onq/2017/08/02/demystifying-3gpp-insiders-perspective-how-4g-and-5g-standards-get-created.

[24] S. E. Hassani, A. Haidine and H. Jebbar, "Road to 5G: Key Enabling Technologies," *Journal of Communications,* vol. 14, no. 11, pp. 1034-1048, 2019.

[25] M. Palmer, "Everything you ever wanted to know about 5G," sifted, 2018. [Online]. Available: https://sifted.eu/articles/5g-mobile-guide/. [Accessed 2021].

[26] P. Delos, B. Broughton and J. Kraft, "Phased Array Antenna Patterns—Part 1: Linear Array Beam Characteristics and Array Factor," analog.com, May 2020. [Online]. Available: https://www.analog.com/en/analog-dialogue/articles/phased-array-antenna-patterns-part1.html. [Accessed 2021].

[27] A. Kaye and D. George, "Transmission of Multiplexed PAM Signals Over Multiple Channel and Diversity Systems," *IEEE Transactions on Communication Technology,* vol. 18, no. 5, pp. 520-526, 1970.

[28] G. Yang and C. Taylor, "4x4 MIMO Boosts 4G and Gives Consumers a Taste of the Gigabit Experience," Huawei, 2017. [Online]. Available: https://www-file.huawei.com/-/media/corporate/pdf/mbb/2020/4x4-mimo.pdf?la=en. [Accessed 2021].

[29] 3GPP, "TS 38.101: User Equipment (UE) radio transmission and reception," 2021.

[30] A. Samuel, "Some Studies in Machine Learning Using the Game of Checkers," *IBM Journal of Research and Development,* vol. 3, pp. 210-229, 1959.

[31] I. Muhammad and Z. Yan, "Supervised Machine Learning Approaches: A Survey," *ICTACT Journal on Soft Computing,* vol. 05, no. 03, pp. 946-952, 2015.

[32] M. Khanum and T. Mahboob, "A Survey on Unsupervised Machine Learning Algorithms for Automation, Classification and Maintenance," *International Journal of Computer Applications ,* vol. 119, no. 13, pp. 34-39, 2015.

[33] M. Ahmed, A. N. Mahmood and J. Hu, "A survey of network anomaly detection techniques," *Journal of Network and Computer Applications,* vol. 60, pp. 19-31, 2016.

[34] R. S. Sutton and A. G. Barto, Reinforcement Learning, Mountain View, CA: Westchester Publishing Services, 2018.

[35] V. Sze, Y.-H. Chen, T.-J. Yang and J. S. Emer, "Efficient Processing of Deep Neural Networks: A Tutorial and Survey," *Proceedings of the IEEE ,* vol. 105, no. 12, pp. 2295-2329, 2017.

[36] I. Goodfellow, Y. Bengio and A. Courville, Deep Learning, MIT Press, 2016.

[37] A. Kathuria, "Intro to optimization in deep learning: Gradient Descent," PaperspaceBlog, 2018. [Online]. Available: https://blog.paperspace.com/intro-to-optimization-in-deep-learning-gradient-descent/. [Accessed 2021].

[38] S. Ruder, "An overview of gradient descent optimization algorithms," *arXiv:1609.04747,* 2016.

[39] D. E. Rumelhart, G. E. Hinton and R. J. Williams, "Learning representations by back-propagating errors," *Nature,* vol. 323, no. Letters to Nature, pp. 533-536, 1986.

[40] C. A. Balanis, Antenna Theory: Analysis and Design, New York: John Wiley & Sons, 1982.

[41] P. Stoica and R. Moses, Introduction to Spectral Analysis, Upper Saddle River, New Jersey: Simon & Schuster, 1997.

[42] M. Bayraktar and G. M. Güvensen, "An Efficient Interference-Aware Constrained Massive MIMO Beamforming for mm-Wave JSDM," *IEEE Access,* vol. 9, pp. 87877-87897, 2021.

[43] I. Chivers and J. Sleightholme, "An Introduction to Algorithms and the Big O Notation," in *Introduction to Programming with Fortran*, Springer, 2015, pp. 359-364.

[44] R. O. Schmidt, "Multiple Emitter Location and Signal Parameter Estimation," *IEEE Trans. Antennas Propagation,* Vols. AP-34, pp. 276-280, 1986.